\documentclass{article}
\usepackage{amsfonts}
\usepackage{amssymb}
\usepackage{spconf,amsmath,epsfig}

\usepackage{hyperref}
\usepackage{algorithmic}
\usepackage{algorithm}
\usepackage{graphicx}
\usepackage{appendix}
\usepackage{color}

\hypersetup{colorlinks=true, linkcolor=blue, citecolor=blue}

\name{Artem Migukin,$^{1\ast }$ Mostafa Agour $^{{2},{3}}$ and Vladimir Katkovnik$^{1}$}

\address{$^1$ \textit{Department of Signal Processing, Tampere University of Technology, FI-33101 Tampere, Finland} \\ $^2$ \textit{Bremer Institut f\"ur angewandte Strahltechnik, Klagenfurter Str. 2, D-28359 Bremen, Germany} \\ $^3$ \textit{Physics Department, Faculty of Science, Aswan University, 81528 Aswan, Egypt} \\ $^*$ \textit{Corresponding author: artem.migukin@tut.fi} \thanks{This work was supported by the Academy of Finland: project no. 138207, 2011-2014. The postgraduate work of Artem Migukin is funded by Tampere Doctoral Programme in Information Science and Engineering (TISE). Mostafa Agour gratefully acknowledge the financial support of the Deutsche Forschungsgemeinschaft (DFG) for funding a part of this work within the frame of the project DynaLiFeS (BE1924/2-1).}}

\input{tcilatex}

\begin{document}

\title{Phase retrieval with background compensation in 4f configuration: advanced augmented Lagrangian technique for amplitude object}
\author{Artem Migukin}
\maketitle

\begin{abstract}
Generally, wave field reconstructions obtained by phase-retrieval algorithms are noisy, blurred and corrupted by various artifacts such as irregular waves, spots, etc. These disturbances, arising due to many factors such as non-idealities of optical system (misalignment, focusing errors), dust on optical elements, reflections, vibration, are hard to be localized and specified. It is assumed that there is a generalized pupil function at the object plane which describes aberrations in the coherent imaging system manifested at the sensor plane. Here we propose a novel two steps phase-retrieval algorithm to compensate these distortions. We first estimate the cumulative disturbance, called \textquotedblleft background\textquotedblright, using special calibration experiments. Then, we use this background for reconstruction of the object amplitude and phase. The second part of the algorithm is based on the maximum likelihood approach and, in this way, targeted on the optimal amplitude and phase reconstruction from noisy data. Numerical experiments demonstrate that the developed algorithm enables the compensation of various typical distortions of the optical track so sharp object imaging for a binary test-chart can be achieved.
\end{abstract}


\section{Introduction}

\noindent The phase contains important information on the shape of the
object, which is useful in metrology and 3D imaging, e.g. microscopy,
astronomy, material analysis, etc. The conventional sensors detect only the
intensity of the light. Since the phase cannot be measured directly and it
is systematically lost in observations, computational phase recovering
techniques are required for imaging and data processing. Phase recovering
and, in general, the reconstruction of the object amplitude and phase is
referred to as the phase-retrieval problem.

Perhaps the first iterative method for phase retrieval from intensity
measurements were the well-known Gerchberg--Saxton algorithm \cite{GS72},
initially for a single observation plane, and its variation devised by
Misell \cite{Missel73} for two defocusing images at different measurement
planes. The idea consisting in the iterative replacement of the estimated
magnitude by measured and prior information is further developed for various
applications by many authors (e.g. \cite{Gonsalves, Fienup78, YG81, YG94}).
Similar methods are proposed for Fresnel instead of Fourier transforms as
the transfer functions of the wave field propagation \cite{Z96, Gureyev3,
Fienup80}. Various phase-retrieval algorithms based on these landmark works
are systematized by Fienup \cite{Fienup82} introducing classical types of
iterative phase-retrieval algorithms. Multiple measurements gain an
observation redundancy that can be exploited in order to improve the quality of the complex-valued object reconstruction \cite{Ivanov,
Pedrini, Almoro-Pedrini}.

The above imaging techniques are mainly based on an \textit{ideal} wave
field propagation modeling derived from the scalar diffraction theory \cite{Goodman}. In practice, wave fields in real coherent
imaging systems and their observations are quite different from those
predicted by theory, hence wave field reconstructions obtained by simulations (i.e. theoretical results) and using real experimental data can dramatically vary. The reconstructions obtained from the real data
differ from simulated ones by multiple and well seen artifacts which can
have a form of disturbed background with irregular waves, spots, random
noise, etc. These systematic distortions appear due to many factors such as
non-idealities of optical system (misalignment, focusing errors, aberrations),
dust on optical elements, reflections, vibration, etc.

\begin{figure}[t!]
\begin{center}
\includegraphics[scale=0.8]{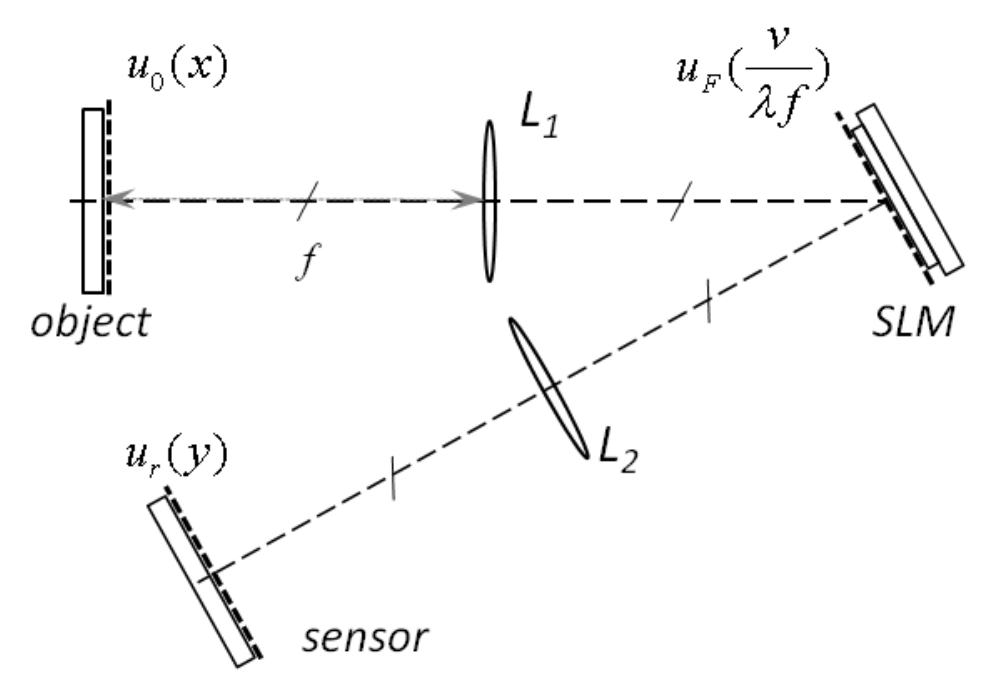}
\end{center}
\caption{\small Experimental setup of the 4f optical system used for recording measurement data \protect\cite{Falldorf10}. The lenses $L_{1}$ and $L_{2}$\ in the 4f configuration provides an accurate mapping of the object wave field to the parallel observation (sensor) plane. An optical mask with the complex-valued transmittance $\mathbf{M}_{r}$ located at the Fourier plane (a phase modulating SLM)\ enables linear filter operations.}
\label{Fig0}
\end{figure}

In this paper we consider a 4f optical system with an SLM located across the
Fourier domain of the first lens. This system is used for capturing multiple
intensity observations at the sensor plane for the phase reconstruction (see
Fig. \ref{Fig0}, \cite{Falldorf10}). The reconstruction from this data is
very sensitive to all disturbing factors because it is an ill-posed problem.
One of the strongest sources of disturbances is the used SLM due to a high
sensitivity of 4f system to modulation of the wave field at the Fourier
plane.

In general, there are diversity of numerical approaches, which are used for calibration 
\cite{Xue}, filtering parasitic reflections \cite{Cuche2000}, compensating
for curvature introduced by microscope objective \cite{Ferraro03, Pedrini01}%
, for aberrations \cite{Zhao12} or astigmatism \cite{Grilli01}. In this work
we follow an essentially different idea. It is assumed that there is a 
\textit{generalized pupil function} \cite[\S 6.4.1]{Goodman} at the object
plane which describes aberrations in the coherent imaging system manifested
at the sensor plane. Namely, we do not try to identify particular sources of
the disturbances but estimate and compensate their accumulated effects by
recalculating them to the entrance pupil of the used 4f configuration. In
the following, the cumulative distortions are referred to as
\textquotedblleft background\textquotedblright\ distortions. Thus, we are
developing two step phase-retrieval algorithm: firstly, we estimate this
background disturbance using special calibration experiments and then use it
to reconstruct the object amplitude and phase. In this work we apply
the variational constrained maximum likelihood formulation with parallel
processing of multiple intensity observations proposed in our previous works 
\cite{MigukinArXiv, VK12-1, VK12-2, MigukinSPIE12}. Moreover, we incorporate
prior information on the true object wave field: in our experiments we
reconstruct a binary object with unknown lower and upper levels.

Let $u_{0}(x)$, $x\in R^{2}$ be a true object wave field at the entrance
pupil of the system. Taking into consideration the non-ideality of the
optical system, we introduce a ``disturbed'' object wave field $\tilde{u}_{0}(x)$
as a product of a typically unknown background (cumulative distortion) wave
field $u_{B}$ by the true object wave field $u_{0}(x)$ as%
\begin{equation}
\tilde{u}_{0}(x)=u_{0}(x)\cdot u_{B}(x)\text{,}  \label{main-model}
\end{equation}%
where the diacritic \ $\tilde{}$\ \ emphasizes the difference of the
corrupted wave field $\tilde{u}_{0}$\ from the ideal one $u_{0}$. The
standard phase-retrieval techniques are able to give the reconstruction of
the disturbed wave field $\tilde{u}_{0}(x)$ only and they are not able to separate
the background in order to estimate the true wave field $u_{0}(x)$.

In our work, we try to reconstruct the disturbances by performing a
calibration procedure (to estimate the background $u_{B}$) and use it to extract the true object $u_{0}$. At first glance, this problem looks
trivial: one may produce the experiments with a known invariant $u_{0}(x)$%
, for instance $u_{0}(x)=1$, obtain the estimate $\hat{u}_{B}(x)$ and then
recalculate the estimate for the object as $\hat{u}_{0}(x)=\tilde{u}_{0}(x)/%
\hat{u}_{B}(x)$. However, a priori information about the object in the used
sparse modeling concerns the true object wave field $u_{0}$ but not the
disturbed one $\tilde{u}_{0}$. Thus, we are processing the recalculated
object estimate\ at each iteration ($\hat{u}_{0}^{t}(x)=\tilde{u}_{0}^{t}(x)/%
\hat{u}_{B}(x),$ $t=0,1,2,...$), and the structure of the developed
iterative phase-retrieval algorithm is therefore essentially different from
the trivial guess.

The paper is organized as follows. In Section \ref{Sec2}, the image
formation in a 4f optical system and the observation model are presented.
The constrained variational approach for the phase retrieval and the sparse
modeling for the object phase and amplitude are introduced in Section \ref%
{Sec3}. The proposed phase-retrieval algorithm with the background
compensation is presented in Section \ref{Sec4}. Numerical experiments for the object wave field reconstructions from real data are shown and discussed in details in Section \ref{Sec5}.

\section{Observation model}

\label{Sec2}\noindent Let us consider the image formation model in a conventional 4f configuration of the coherent imaging system linking complex amplitudes at the object and measurement planes.

Let us denote complex amplitudes at the object and measurement (sensor)
planes by $u_{0}(x)$ and $u_{r}(y)$, respectively. The lenses $L_{1}$ and $%
L_{2}$\ with the focal length $f$ arranged in the 4f configuration provides
an accurate mapping of the object wave field into the parallel measurement
plane. A reflective phase\textbf{\ }modulating spatial light modulator (SLM)
is placed at the Fourier plane of the first lens \cite{Falldorf10}. The used
4f optical system is illustrated in Fig. \ref{Fig0}.

Let us assume for a moment that there are no distortions in the optical
track. It is well known that the link between the wave fields at the object $%
u_{0}(x)$ and the Fourier planes $u_{F}(\frac{v}{\lambda f})$ is given as
follows \cite{Goodman} 
\begin{equation}
u_{F}(\frac{v}{\lambda f})=\frac{1}{i\lambda f}\mathcal{F}\{u_{0}(x)\}(\frac{%
v}{\lambda f}),  \label{Eq01}
\end{equation}%
where $\mathcal{F}\{\cdot \}$ denotes the 2D integral Fourier transform, $%
\lambda $ is a wavelength.

If the optical mask (SLM) inserted at the Fourier plane has the
complex-valued transmittance $M_{r}(\frac{v}{\lambda f})$, then the output
of the optical system is defined as 
\begin{equation}
u_{r}(y)=\frac{1}{i\lambda f}\mathcal{F}\{u_{F}(\frac{v}{\lambda f})\cdot
M_{r}(\frac{v}{\lambda f})\}(-y).  \label{Eq02}
\end{equation}

All these wave field distributions are given in the 2D lateral coordinates:
here we use the variables $x$, $y$, $v\in 
\mathbb{R}
^{2}$\ for the object, sensor and Fourier planes, respectively.

\subsection{Discrete modeling}

For discrete modeling, the continuous arguments are replaced by the digital ones with a corresponding replacement of the continuous functions by their discrete counterparts: $u_{0}(x)\rightarrow u_{0}(k_{1}\Delta_{x_{1}},k_{2}\Delta _{x_{2}})$, $u_{r}(y)\rightarrow u_{r}(l_{1}\Delta_{y_{1}},l_{2}\Delta _{y_{2}})$, $u_{F}(\frac{v}{\lambda f})\rightarrow
u_{F}(\frac{\Delta _{v_{1}}}{\lambda f}\eta _{1},\frac{\Delta _{v_{2}}}{\lambda f}\eta _{2})$\ with $2D$ integer arguments $k=(k_{1},k_{2}),$ $l=(l_{1},l_{2})$ and $\eta =(\eta _{1},\eta _{2})$. This discretization is dictated by the use of a digital camera and a pixelated SLM as a 2D array of liquid crystal cells. Thus, we hereafter consider the discrete wave fields at the object $u_{0}[k_{1},k_{2}]$, Fourier $u_{F}[\eta _{1},\eta _{2}]$ and sensor planes $u_{r}[l_{1},l_{2}]$ with various pixel sizes $\Delta_{x_{1}}\times \Delta _{x_{1}}$, $\Delta _{v_{1}}\times \Delta _{v_{1}}$\ and $\Delta _{y_{1}}\times \Delta _{y_{1}}$, respectively. In general, these images can be rectangular of different size $N_{x_{1}}\times N_{x_{1}}$, $N_{v_{1}}\times N_{v_{2}}$\ and $N_{y_{1}}\times N_{y_{1}}$, respectively.

We use a vector-matrix notation for complex-valued distributions of the wave fields. 2D discrete distributions (matrices) are vectorized to the complex-valued column vector \cite{Jan}. Bold lower case characters are used for the vectors. Matrices are defined by bold upper case to distinguish them from vectors. Thus, $\mathbf{u}_{0}[k]$, $\mathbf{u}_{F}[\eta ]$\ and $\mathbf{u}_{r}[l]$\ are column vectors constructed by vectorization of the corresponding 2D discrete wave field distributions at the object $\mathbf{U}_{0}[k_{1},k_{2}]$, Fourier $\mathbf{U}_{F}[\eta _{1},\eta _{2}]$\ and sensor planes $\mathbf{U}_{r}[l_{1},l_{2}]$, respectively.

In this work we assume that the pixel size at the object and sensor planes is the same ($\Delta _{x_{1}}=\Delta _{y_{1}}$, $\Delta _{x_{2}}=\Delta _{y_{2}}$)\ and images $\mathbf{U}_{0}$, $\mathbf{U}_{F}$ and\ $\mathbf{U}_{r}$\ are of the same size $N_{x}\times N_{y}$ for all planes. Let us also assume that the following conditions are fulfilled \cite{Kreis}%
\begin{equation}
\Delta _{v_{1}}\Delta _{x_{1}}N_{x}=\lambda f\text{, }\Delta _{v_{2}}\Delta
_{x_{2}}N_{y}=\lambda f  \label{sc1}
\end{equation}

Then, discretization of the integral in Eq. (\ref{Eq01}) defines $\mathbf{U}%
_{F}[\eta _{1},\eta _{2}]$ as 2D discrete Fourier transform (DFT) of $%
\mathbf{U}_{0}[k_{1},k_{2}]$ in the form%
\begin{eqnarray}
&&\mathbf{U}_{F}[\eta _{1},\eta _{2}]=  \label{F01} \\
&&\frac{\Delta _{x_{1}}\Delta _{x_{2}}}{i\lambda f}\underset{k_{1}=\frac{%
-N_{x_{1}}}{2}}{\overset{\frac{N_{x_{1}}}{2}-1}{\sum }}\underset{k_{2}=\frac{%
-N_{x_{2}}}{2}}{\overset{\frac{N_{x_{2}}}{2}-1}{\sum }}\mathbf{U}%
_{0}(k_{1}\Delta _{x_{1}},k_{2}\Delta _{x_{2}})\times  \notag \\
&&\times \exp (\frac{2\pi }{i\lambda f}\cdot (\eta _{1}\Delta
_{v_{1}}k_{1}\Delta _{x_{1}}+\eta _{2}\Delta _{v_{2}}k_{2}\Delta _{x_{2}})) \notag \\
&&\frac{-N_{v_{1}}}{2}\leq \eta _{1}\leq \frac{N_{v_{1}}}{2}-1,\text{ }\frac{-N_{v_{2}}}{2}\leq \eta _{2}\leq \frac{N_{v_{2}}}{2}-1 
\notag
\end{eqnarray}

Similarly, discrete model for Eq.(\ref{Eq02}) has the form 
\begin{equation}
\mathbf{U}_{r}[l_{1},l_{2}]=\frac{\Delta _{v_{1}}\Delta _{v_{2}}}{i\lambda f}F\{\mathbf{U}_{F}\circ \mathbf{M}_{r}\}[-l_{1},-l_{2}]   \label{F02}
\end{equation}%
where $F\{\cdot \}$ in Eqs. (\ref{F01}) and (\ref{F02}) denotes the 2D DFT
operator. $\circ $\ stands for the Hadamard (elementwise) product and $%
\mathbf{M}_{r}[\eta _{1},\eta _{2}]$\ is the discretized optical mask at the
Fourier plane. Note that the wave field propagation can be easily realized
much faster using FFT (cf. \cite{VK12-2, MigukinSPIE12}).

Taking into account the vector-matrix notation and the distortions in the
real optical system (Eq. (\ref{main-model})), the forward wave field
propagation from the object to the sensor plane can be given in the form 
\begin{equation}
\mathbf{u}_{r}=\mathbf{A}_{r}\cdot \mathbf{\tilde{u}}_{0}\text{, }r=1,...K,
\label{WFPM}
\end{equation}%
where $\mathbf{\tilde{u}}_{0}\in\mathbb{C}^{n\times 1}$ is a complex-valued vector, corresponding to
the disturbed object discrete 2D wave field distribution, $n=N_{x}\cdot N_{y}$. $\mathbf{A}_{r}\in 
\mathbb{C}
^{n\times n}$ is a forward propagation operator corresponding to the optical
mask $\mathbf{M}_{r}$\ at the Fourier plane (programmed using the SLM), and $K$ is a
number of these various optical masks.

\subsection{Noisy intensity observations}

Assume that we have a set of $K$\ experiments produced with different masks $%
\{\mathbf{M}_{r}\}_r$, $r=1,...,K$. The problem is to reconstruct a
complex-valued true object wave field $\mathbf{u}_{0}$\ from multiple noisy
intensity observations $\{\mathbf{o}_{r}\}$\ measured at the sensor plane.
These measurements are represented in the vector-matrix notation following
to Eq. (\ref{WFPM})\ as follows 
\begin{equation}
\mathbf{o}_{r}[l]=|\mathbf{u}_{r}[l]|^{2}+\mathbf{\varepsilon }_{r}[l],\text{
}r=1,...K,  \label{observation_1}
\end{equation}%
where the noise is assumed to be\ zero-mean Gaussian with the variance $%
\sigma _{r}^{2}$, $\mathbf{\varepsilon }_{r}[l]\sim \mathcal{N}(0,\sigma
_{r}^{2})$, independent for different $l$ and $r$. The observation vectors $%
\{\mathbf{o}_{r}\}$ correspond to the 2D distributions on the regular
discrete grid located at the sensor plane.

\section{Sparse object modeling and variational formulation}

\label{Sec3}\noindent It is assumed in sparse modeling approach that the
``true'' object distribution $\mathbf{u}_{0}$ can be approximated by a small
number of non-zero elements of basis functions. The ideal basis functions
for the object approximation are unknown a priori and selected from a given
set of potential bases (dictionaries). In general, we deal with a complex-valued object wave field,
and hence consider nonlinear modeling of the wave field with separate
approximations for the object phase and amplitude \cite{MigukinArXiv,
VK12-1, VK12-2, MigukinSPIE12}. We represent the object wave field in the
form $\mathbf{u}_{0}=\mathbf{a}_{0}\circ \exp (j\cdot \mathbf{\varphi }_{0})$%
, where $\mathbf{a}_{0}\triangleq abs(\mathbf{u}_{0})\in 
\mathbb{R}
^{n}$ and $\mathbf{\varphi }_{0}\triangleq angle(\mathbf{u}_{0})\in 
\mathbb{R}
^{n}$ denote the object amplitude and phase, respectively. Sparse object
approximation can be given in the analysis or synthesis form as follows%
\begin{equation}
\begin{array}{cc}
\mathbf{\theta }_{a}=\mathbf{\Phi }_{a}\cdot abs(\mathbf{u}_{0}),\mathbf{%
\theta }_{\varphi }=\mathbf{\Phi }_{\varphi }\cdot angle(\mathbf{u}_{0}) & 
\text{(analysis)} \\ 
\mathbf{a}_{0}=\mathbf{\Psi }_{a}\cdot \mathbf{\theta }_{a},\mathbf{\varphi }%
_{0}=\mathbf{\Psi }_{\varphi }\cdot \mathbf{\theta }_{\varphi } & \text{%
(synthesis)}%
\end{array}
\label{Eq07}
\end{equation}

Here $\mathbf{\Psi }_{a}$,$\mathbf{\Psi }_{\varphi }$ and $\mathbf{\Phi }%
_{a} $, $\mathbf{\Phi }_{\varphi }$ are the frame transform matrices, and
the vector $\mathbf{\theta }_{a}$, $\mathbf{\theta }_{\varphi }\in 
\mathbb{R}
^{m}$ can be considered as a spectrum ($m\gg n$) in a parametric data
adaptive approximation. Subindices $a$ and $\varphi $\ are shown for the
amplitude and phase, respectively. It is recognized that, in contrast to
classical orthonormal bases ($m=n$), overcomplete frame based modeling is a
much more efficient for imaging \cite{Elad, Han} and results in a better
wave field reconstruction accuracy. The sparsity of approximation is
characterized by either the $\ell _{0}$ norm $||\mathbf{\theta }||_{0}$
defined as a number of non-zero components of the vector $\mathbf{\theta }$
or the $\ell _{1}$ norm as a sum of absolute values of components of the
vector $||\mathbf{\theta }||_{1}=\sum_{s}|\mathbf{\theta }_{s}|$. A smaller
value of the norm means a higher sparsity of approximation. Note that
results obtained by $\ell _{0}$ or $\ell _{1}$ norms are shown to be closed
to each other \cite{Donoho}, what allows replacing the nonconvex $\ell _{0}$%
\ norm by the convex $\ell _{1}$ norm in many variational settings.

The main intention is to find sparsest (shortest) models for phase and
amplitude with smallest values of the $\ell _{0}$ or $\ell _{1}$ norms. The
separate sparse modeling for the object phase and amplitude is realized via
the powerful BM3D-frame filter, specified for denoising and other imaging
problems \cite{VK2011, Aram, Aram12}.

Assume that the background wave field $\mathbf{u}_{B}$ is given. Taking into
account the sparse modeling for the object amplitude and phase, the wave
field reconstruction is performed by minimization of the criterion $\mathcal{%
J}$%
\begin{gather}
\mathcal{J}=\sum_{r=1}^{K}\frac{1}{2\sigma ^{2}}||\mathbf{o}_{r}-|\mathbf{u}%
_{r}|^{2}||_{2}^{2}+\tau _{a} ||\mathbf{\theta }_{a}||_{p}+\tau
_{\varphi } ||\mathbf{\theta }_{_{\varphi }}||_{p}  \label{criterionJ}
\\
\text{subject to }\mathbf{u}_{r}=\mathbf{A}_{r}\cdot \mathbf{\tilde{u}}_{0}%
\text{, }r=1,...K\text{,}  \label{criterionJa} \\
\mathbf{\tilde{u}}_{0}=\mathbf{u}_{0}\circ \mathbf{u}_{B}
\label{criterionJb} \\
\mathbf{\theta }_{a}=\mathbf{\Phi }_{a}\cdot abs(\mathbf{u}_{0}),\mathbf{%
\theta }_{\varphi }=\mathbf{\Phi }_{\varphi }\cdot angle(\mathbf{u}_{0})%
\text{,}  \label{criterionJc} \\
\mathbf{a}_{0}=\mathbf{\Psi }_{a}\cdot \mathbf{\theta }_{a},\mathbf{\varphi }%
_{0}=\mathbf{\Psi }_{\varphi }\cdot \mathbf{\theta }_{\varphi }
\label{criterionJd}
\end{gather}%
where $||\mathbf{\cdot }||_{2}$ stands for the Euclidean norm and
regularization terms for phase and amplitude are taken using the $\ell _{p}$
norms ($p=\{0,1\}$). The positive parameters $\tau _{a}$ and $\tau _{\varphi
}$ \ in Eq. (\ref{criterionJ}) define a balance between the fit of
observations, smoothness of the wave field reconstruction and the complexity
of the used model (cardinality of spectra $\mathbf{\theta }_{a}$ , $\mathbf{%
\theta }_{\varphi }$ of the object amplitude and phase). Note that the
constraint for the forward wave field propagation (\ref{criterionJa}) is
presented for the disturbed object wave field $\mathbf{\tilde{u}}_{0}$ (Eq. (%
\ref{criterionJb})), and the used sparse modeling is given for the true
object: namely, the analysis (\ref{criterionJc}) and synthesis (\ref%
{criterionJd}) are calculated using the frame transform matrices for the
compensated object $\mathbf{u}_{0}[k]=\mathbf{\tilde{u}}_{0}[k]/\mathbf{u}%
_{B}[k]$ (rather for the object amplitude and phase).

\subsection{Multi-objective optimization}

It is shown in \cite{Aram12} that a multi-objective optimization can be much
more efficient than the minimization of the single criterion $\mathcal{J}$
due to a simpler implementation (filtering and inverse procedure are
decoupled) and resulting better reconstruction quality. Thus, instead of the
constrained minimization of (\ref{criterionJ}) we arrive at the
unconstrained minimization of two criterion functions $\mathcal{J}_{1}$ and $%
\mathcal{J}_{2}$\ with changing the constraints for sparse modeling by the
quadratic penalties with positive weights 
\begin{eqnarray}
\mathcal{J}_{1}=\frac{1}{\gamma _{0}}||\mathbf{\tilde{u}}_{0}-\mathbf{\tilde{v}}_{0}||_{2}^{2}+ \label{criterionJ1} \\
+\sum_{r=1}^{K}\frac{1}{\sigma ^{2}}[\frac{1}{2}||\mathbf{o}_{r}-|\mathbf{u}_{r}|^{2}||_{2}^{2}+\frac{1}{\gamma _{r}}||\mathbf{u}_{r}-\mathbf{A}_{r}\cdot \mathbf{\tilde{u}}_{0}||_{2}^{2}+ \notag \\ +\frac{2}{\gamma _{r}}\func{Re}\{\mathbf{\Lambda }_{r}^{H}\cdot (\mathbf{u}_{r}-\mathbf{A}_{r}\cdot \mathbf{\tilde{u}}_{0})\},  \notag
\end{eqnarray}
\begin{eqnarray}
\mathcal{J}_{2}=\tau _{a}||\mathbf{\theta }_{a}||_{p}+\frac{1}{%
2\gamma _{a}}||\mathbf{\theta }_{a}-\mathbf{\Phi }_{a}\cdot abs(\mathbf{u}%
_{0})||_{2}^{2}+  \label{criterionJ2} \\
+\tau _{_{\varphi }}||\mathbf{\theta }_{\varphi }||_{p}+\frac{1}{%
2\gamma _{\varphi }}||\mathbf{\theta }_{\varphi }-\mathbf{\Phi }_{\varphi
}\cdot angle(\mathbf{u}_{0})||_{2}^{2},  \notag
\end{eqnarray}%
where $(\cdot )^{H}$ stands in Eq. (\ref{criterionJ1}) for the Hermitian
conjugate, $\mathbf{v}_{0}=\mathbf{\Psi }_{a}\mathbf{\theta }_{a}\circ \exp
(j\cdot \mathbf{\Psi }_{\varphi }\mathbf{\theta }_{\varphi })$\ is an
approximation of the complex-valued object distribution $\mathbf{u}_{0}$, $%
\mathbf{\tilde{v}}_{0}=\mathbf{v}_{0}\circ \mathbf{\mathbf{u}}_{B}$. $\{%
\mathbf{\Lambda }_{r}\}\in 
\mathbb{C}
^{n}$\ are the complex-valued vectors of the Lagrange multipliers (see \cite{MigukinAL}).
Note that the linear and quadratic penalties related to the forward
propagation are involved with the same positive parameters $\gamma _{r}$.
The analysis and synthesis constraints in Eqs. (\ref{criterionJc}) and (\ref%
{criterionJd}) are replaced by the penalties with the corresponding positive
parameters $\gamma _{a}$ , $\gamma _{\varphi }$ and $\gamma _{0}$, in Eqs. (\ref{criterionJ1}) and (\ref{criterionJ2}), what is a standard tools to deal
with constrained optimization \cite{Bertsekas}.

Note that the criterion function $\mathcal{J}_{2}$ is separable with respect
to $\mathbf{\theta }_{a}$ and $\mathbf{\theta }_{\varphi }$, thus it can be
rewritten as $\mathcal{J}_{2}=\mathcal{J}_{2,a}+\mathcal{J}_{2,\varphi }$,
where%
\begin{eqnarray}
&&\mathcal{J}_{2,a}(\mathbf{\theta }_{a},abs(\mathbf{u}_{0}))=
\label{criterionJ2a} \\
&=&\tau _{a}\cdot ||\mathbf{\theta }_{a}||_{p}+\frac{1}{2\gamma _{a}}||%
\mathbf{\theta }_{a}-\mathbf{\Phi }_{a}\cdot abs(\mathbf{u}_{0})||_{2}^{2} 
\notag \\
&&\mathcal{J}_{2,\varphi }(\mathbf{\theta }_{\varphi },angle(\mathbf{u}%
_{0}))=  \label{criterionJ2b} \\
&=&\tau _{_{\varphi }}\cdot ||\mathbf{\theta }_{\varphi }||_{p}+\frac{1}{%
2\gamma _{\varphi }}||\mathbf{\theta }_{\varphi }-\mathbf{\Phi }_{\varphi
}\cdot angle(\mathbf{u}_{0})||_{2}^{2}  \notag
\end{eqnarray}

\begin{figure}[b!]
\begin{center}
\includegraphics[scale=0.48]{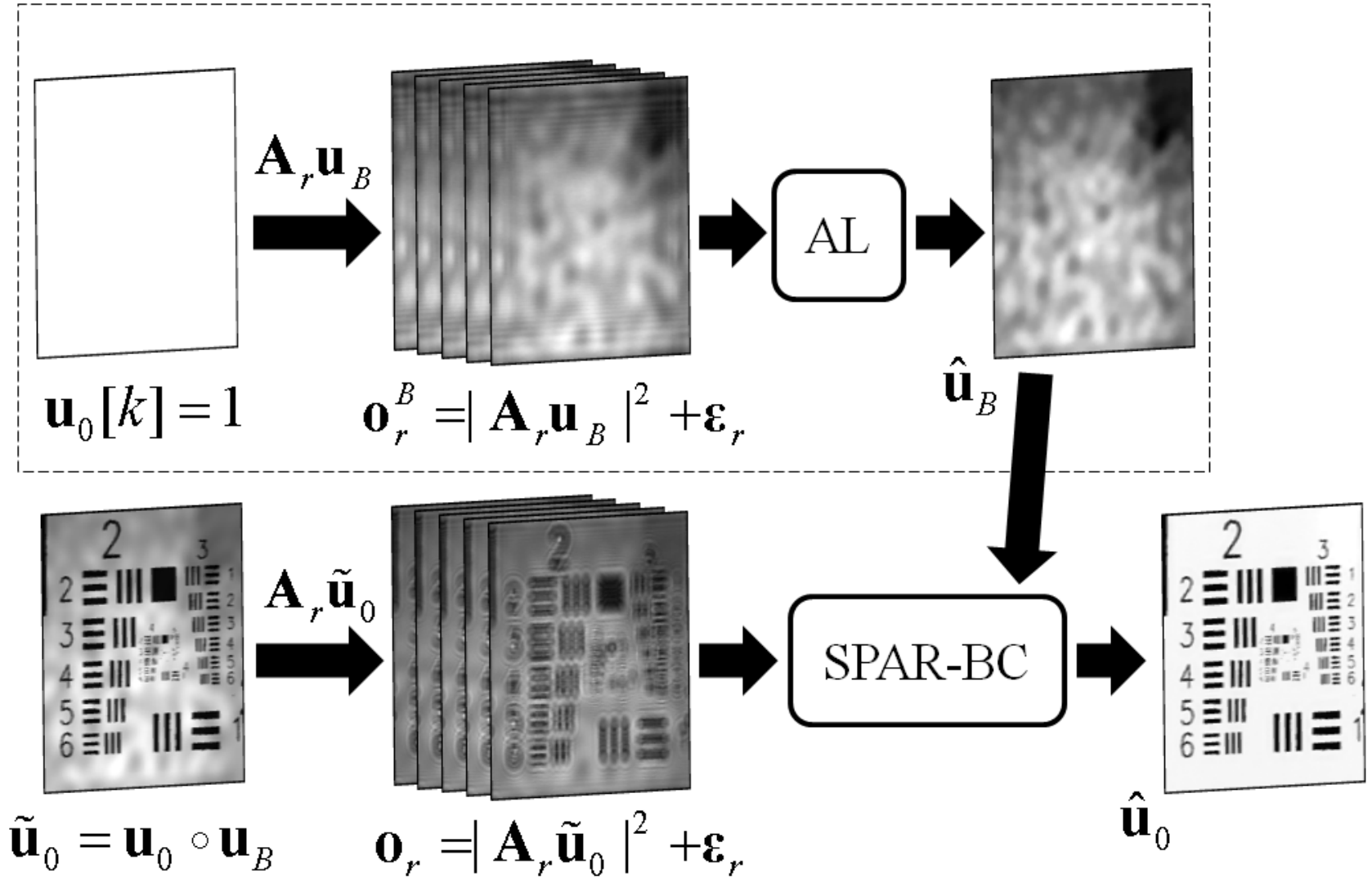}
\end{center}
\caption{ A flowchart of the two steps phase retrieval technique with the background compensation. The upper block highlighted by a dashed line represents the background
calibration procedure, where the complex-valued estimate of $\mathbf{u}_{B}$ is found by $AL$ \protect\cite{MigukinAL}. The reconstruction of the object using the background estimate\ is obtained by the proposed $SPAR-BC$ method. }
\label{Fig1}
\end{figure}

In contrast to \cite{VK12-2}, where only a quadratic penalization is used
for the forward propagation model $\mathbf{u}_{r}=\mathbf{A}_{r}\cdot 
\mathbf{\tilde{u}}_{0}$, here we use the linear Lagrangian summand $\func{Re}%
\{\mathbf{\Lambda }_{r}^{H}\cdot (\mathbf{u}_{r}-\mathbf{A}_{r}\cdot \mathbf{%
\tilde{u}}_{0})\}$ in Eq.(\ref{criterionJ1}). Thus, $\mathcal{J}_{1}$
becomes the augmented Lagrangian objective function at least with respect to
this particular constraint.

\section{Proposed algorithm}

\label{Sec4}\noindent Following our assumption that the object wave field
is degraded by the distortions accumulated in the background wave field $%
\mathbf{u}_{B}$ ($\mathbf{\tilde{u}}_{0}=\mathbf{u}_{0}\circ \mathbf{\mathbf{%
u}}_{B}$, cf. Eq. (\ref{main-model})), special calibration experiments are
produced in order to estimate these disturbances. Thus, it is assumed that
two sets of experiments are made consistently under the same conditions.
These experiments result in two sets of observations $\{\mathbf{o}_{r}^{B}\}$
and $\{\mathbf{o}_{r}\}$\ used to estimate the background $\mathbf{\hat{u}}%
_{B}$ and the the object $\mathbf{u}_{0}$, respectively. The flowchart of
the proposed two steps algorithm is shown in Fig. \ref{Fig1}.

\subsection{Background reconstruction}

Firstly, we record a number of intensity observations $\{\mathbf{o}%
_{r}^{B}\} $ corresponding to various optical masks $\{\mathbf{M}_{r}\}$ for
the free space object (test-image $\mathbf{u}_{0}[k]=1$) and find a
complex-valued estimate $\mathbf{\hat{u}}_{B}$ ($\mathbf{\tilde{u}}%
_{0}\equiv \mathbf{u}_{B} $)\ by optimization of the criterion function 
\begin{eqnarray}
&&\mathcal{J}_{AL} = \mu ||\mathbf{u}_{B}||_{2}^{2} + \sum_{r=1}^{K}\frac{1}{\sigma ^{2}}[\frac{1}{2}||\mathbf{o}_{r}^{B}-|\mathbf{u}_{r}|^{2}||_{2}^{2}+ \label{AL0}\\ &&+\frac{1}{\gamma _{r}}||\mathbf{u}_{r}-\mathbf{A}_{r} \mathbf{u}_{B}||_{2}^{2}+\frac{2}{\gamma _{r}}\func{Re}\{\mathbf{\Lambda }_{r}^{H} (\mathbf{u}_{r}-\mathbf{A}_{r} \mathbf{u}_{B})\}  \notag
\end{eqnarray}%
as it is described in \cite{MigukinAL}. The main difference of the structure
of $\mathcal{J}_{AL}$\ from $\mathcal{J}_{1}$\ consists in the last
quadratic penalty term. $\mu $ in Eq. (\ref{AL0}) is the Tikhonov
regularization parameter which defines a balance between the prior
information on $\mathbf{u}_{B}$ and the fitting\ of calculated intensities $|%
\mathbf{u}_{r}|^{2}$ to the given observations $\mathbf{o}_{r}^{B}$. The
estimate $\mathbf{\hat{u}}_{B}$\ can be computed using the following
algorithm%
\begin{eqnarray}
&&\text{Repeat for }t=0,1,2,...  \label{AL1} \\
&&\text{Repeat for }r=1,...K  \notag \\
&&\mathbf{u}_{r}^{t}=\arg \min_{\mathbf{u}_{r}}\mathcal{J}_{AL}(\mathbf{o}%
_{r}^{B},\mathbf{u}_{B}^{t},\mathbf{u}_{r},\mathbf{\Lambda }_{r}^{t},) 
\notag \\
&&\mathbf{\Lambda }_{r}^{t+1}=\mathbf{\Lambda }_{r}^{t}+\alpha _{r}\cdot (%
\mathbf{u}_{r}^{t}-\mathbf{A}_{r}\cdot \mathbf{u}_{B}^{t})\text{ }  \notag \\
&&\text{End on }r  \notag \\
&&\mathbf{u}_{B}^{t+1}=\arg \min_{\mathbf{u}_{B}}\mathcal{J}_{AL}(\{\mathbf{o%
}_{r}^{B}\},\mathbf{u}_{B},\{\mathbf{u}_{r}^{t}\},\{\mathbf{\Lambda }%
_{r}^{t}\}),  \notag \\
&&\text{End on }t  \notag
\end{eqnarray}

This algorithm without background compensation and additional object
filtering is identical to the $AL$ algorithm originated in \cite{MigukinAL}
but presented with respect to the background $\mathbf{u}_{B}$.

This stage is shown in the upper block of Fig. \ref{Fig1} highlighted by a
dashed line.

\subsection{Object reconstruction}

Secondly, we record intensity measurements for an object $\{\mathbf{o}_{r}\}$
using the same optical masks $\{\mathbf{M}_{r}\}$ as before and reconstruct
the true object wave field using the found background estimate $\mathbf{\hat{%
u}}_{B}$.

According to the general idea of the multi-objective optimization, where the
alternating minimization of $\mathcal{J}_{1}$ with respect to $\mathbf{%
\tilde{u}}_{0}=\mathbf{u}_{0}\circ \mathbf{u}_{B}$, $\{\mathbf{u}_{r}\}$ and
minimization of $\mathcal{J}_{2}$ with respect to $\mathbf{\theta }_{a},%
\mathbf{\theta }_{\varphi }$ are used \cite{Aram, Aram12, VK12-2}, we arrive
at the following iterative algorithm%
\begin{eqnarray}
\mathbf{u}_{0}^{t}[k] &=&\mathbf{\tilde{u}}_{0}^{t}[k]/\mathbf{\hat{u}}%
_{B}[k]  \label{opt3a} \\
\mathbf{\theta }_{a}^{t} &=&\arg \min_{\mathbf{\theta }_{a}}\mathcal{J}%
_{2,a}(\mathbf{\theta }_{a},abs(\mathbf{u}_{0}^{t}))  \label{opt2a1} \\
\mathbf{\theta }_{\varphi }^{t} &=&\arg \min_{\mathbf{\theta }_{\varphi }}%
\mathcal{J}_{2,\varphi }(\mathbf{\theta }_{\varphi },angle(\mathbf{u}%
_{0}^{t}))  \label{opt2a2} \\
\mathbf{\tilde{v}}_{0}^{t} &=&\mathbf{\Psi }_{a}\mathbf{\theta }%
_{a}^{t}\circ \exp (j\cdot \mathbf{\Psi }_{\varphi }\mathbf{\theta }%
_{\varphi }^{t})\circ \mathbf{\mathbf{\hat{u}}}_{B}  \label{opt2b3} \\
\mathbf{u}_{r}^{t} &=&\arg \min_{\mathbf{u}_{r}}\mathcal{J}_{1}(\mathbf{o}%
_{r},\mathbf{\tilde{u}}_{0}^{t},\mathbf{u}_{r},\mathbf{\Lambda }_{r}^{t},%
\mathbf{\tilde{v}}_{0}^{t}),  \label{opt1} \\
\mathbf{\Lambda }_{r}^{t+1} &=&\mathbf{\Lambda }_{r}^{t}+\alpha _{r}\cdot (%
\mathbf{u}_{r}^{t}-\mathbf{A}_{r}\cdot \mathbf{\tilde{u}}_{0}^{t}),\text{ }%
r=1,...K,  \label{opt10} \\
\mathbf{\tilde{u}}_{0}^{t+1} &=&\arg \min_{\mathbf{\tilde{u}}_{o}}\mathcal{J}%
_{1}(\{\mathbf{o}_{r}\},\mathbf{\tilde{u}}_{0},\{\mathbf{u}_{r}^{t}\},\{%
\mathbf{\Lambda }_{r}^{t}\},\mathbf{\tilde{v}}_{0}^{t})  \label{opt4}
\end{eqnarray}

Here we first estimate the true object (Eq.(\ref{opt3a})) by
compensation of the disturbed $\mathbf{\tilde{u}}_{0}^{t}$ with the
background estimate $\mathbf{\hat{u}}_{B}$. It results in the object
amplitude and phase estimates. Then, Eqs. (\ref{opt2a1}) and (\ref{opt2a2})
enable the spectrum estimates of the object amplitude and phase by
thresholding in the BM3D-frame domain with the thresholds $\tau _{a}\gamma
_{a}$\ and $\tau _{\varphi }\gamma _{\varphi }$, respectively \cite{VK2011,
Aram}. Eq. (\ref{opt2b3}) corresponds to the synthesis of the approximation $%
\mathbf{\tilde{v}}_{0}^{t}$ of the disturbed object from the calculated
spectra for the true object amplitude and phase and using the background $%
\mathbf{\hat{u}}_{B}$. Together the operations in Eqs. (\ref{opt2a1})--(\ref%
{opt2b3}) related to the optimization of $\mathcal{J}_{2}$ can be rewritten
in a more compact form as follows%
\begin{eqnarray}
&&\mathbf{a}_{0}^{t+1/2}=BM3D_{a}(abs(\mathbf{u}_{0}^{t})),  \label{Eq03} \\
&&\mathbf{\varphi }_{0}^{t+1/2}=BM3D_{\varphi }(angle(\mathbf{u}_{0}^{t})), 
\notag \\
&&\mathbf{\tilde{v}}_{0}^{t}=\mathbf{a}_{0}^{t+1/2}\circ \exp (j\cdot 
\mathbf{\varphi }_{0}^{t+1/2})\circ \mathbf{\mathbf{\hat{u}}}_{B}  \notag
\end{eqnarray}%
where $BM3D(\cdot )$\ denotes hereafter the processing by the BM3D filter,
and the corresponding subindices $a$ and $\varphi $ emphasize that the
filtering is performed with different parameters and different transform
matrices $\mathbf{\Psi }_{\cdot }$ and $\mathbf{\Phi }_{\cdot }$ for the
amplitude and phase, respectively. In our implementation the analysis and
synthesis operations, the thresholding and calculation of the matrices $%
\mathbf{\Psi }_{\cdot }$ and $\mathbf{\Phi }_{\cdot }$ are integrated in a
single block which is called $BM3D$\ filter.

Eqs. (\ref{opt1}), (\ref{opt4}) are the optimization steps for $\mathcal{J}%
_{1}$: the computation of the complex-valued wave field estimates $\{\mathbf{%
u}_{r}\}$ at the sensor planes and the disturbed object $\mathbf{\tilde{u}}%
_{0}$ from noisy intensity observations $\{\mathbf{o}_{r}\}$. The update of
the Lagrange variables $\mathbf{\Lambda }_{r}$ is shown in Eq. (\ref{opt10}). This second stage is illustrated in Fig. \ref{Fig1} under the mentioned
block for the background estimation.

\subsection{Sparse Phase Amplitude Reconstruction with Background Compensation ($SPAR-BC$)}
\label{SecSPARBC}

Following the mentioned two main steps --\ Eqs. (\ref{AL1})
for the background $\mathbf{\hat{u}}_{B}$\ estimation and Eqs. (\ref{opt3a}%
)--(\ref{opt4}) for the true complex-valued object extraction -- we
formulate the advanced phase-retrieval approach with background
compensation. Taking into account Eqs. (\ref{Eq03}), the reconstruction of
the true object wave field is performed by the proposed iterative algorithm called \textbf{S}parse \textbf{P}hase \textbf{A}mplitude \textbf{R}econstruction with \textbf{B}ackground \textbf{C}ompensation ($SPAR-BC$).

\begin{table}[tb!]
\begin{tabular}{c}
\hline
\textbf{Algorithm}: $SPAR-BC$ \\ 
\textbf{Input}: $\{\mathbf{o}_{r}^{B}\}_{r=1}^{K}$, \ $\{\mathbf{o}%
_{r}\}_{r=1}^{K}$ \\ 
\textbf{Initialization}: $\mathbf{\hat{u}}_{B}$, \ $\mathbf{\tilde{u}}%
_{0}^{0}$, \ $\{\mathbf{\Lambda }_{r}^{0}\}$ \\ 
\textbf{Repeat for} $t=0,1,2...$ \\ 
1. Object update (background compensation): \\ 
$\mathbf{u}_{0}^{t}[k]=\mathbf{\tilde{u}}_{0}^{t}[k]/\mathbf{\hat{u}}_{B}[k]$
\\ 
2. BM3D filtering: \\ 
$\mathbf{a}_{0}^{t+1/2}=BM3D_{a}(abs(\mathbf{u}_{0}^{t}))$, \\ 
$\mathbf{\varphi }_{0}^{t+1/2}=BM3D_{\varphi }(angle(\mathbf{u}_{0}^{t}))$
\\ 
3. Object approximation synthesis: \\ 
$\mathbf{v}_{0}^{t+1}=\mathbf{a}_{0}^{t+1/2}\circ \exp (j\cdot \mathbf{%
\varphi }_{0}^{t+1/2})$ \\ 
\textbf{Repeat for} $r=1,...K$ \\ 
4. Forward propagation: \\ 
$\mathbf{u}_{r}^{t+1/2}=\mathbf{A}_{r}\cdot \mathbf{\tilde{u}}_{0}^{t}$ \\ 
5. Fitting to observations: \\ 
$\mathbf{u}_{r}^{t+1}[l]=\mathcal{G}(\mathbf{o}_{r}[l],\mathbf{u}%
_{r}^{t+1/2}[l],\mathbf{\Lambda }_{r}^{t}[l])$ $\forall l$ \\ 
6. Lagrange multipliers update: \\ 
$\mathbf{\Lambda }_{r}^{t+1}=\mathbf{\Lambda }_{r}^{t}+\alpha _{r}\cdot (%
\mathbf{u}_{r}^{t+1}-\mathbf{u}_{r}^{t+1/2})$ \\ 
\textbf{End on} $r$ \\ 
7. Disturbed object update: \\ 
$\mathbf{\tilde{u}}_{0}^{t+1}=(\sum_{r=1}^{K}\frac{1}{\gamma _{r}\sigma
_{r}^{2}}\mathbf{A}_{r}^{H}\mathbf{A}_{r}+\frac{1}{\gamma _{0}}\cdot \mathbf{%
I}_{n\times n})^{-1}\times $ \\ 
$\times \sum_{r=1}^{K}\frac{1}{\gamma _{r}\sigma _{r}^{2}}\mathbf{A}%
_{r}^{H}\cdot (\mathbf{u}_{r}^{t+1}+\mathbf{\Lambda }_{r}^{t})+\frac{1}{%
\gamma _{0}}\cdot (\mathbf{\hat{u}}_{B}\circ \mathbf{v}_{0}^{t+1})$ \\ 
\textbf{End on} $t$ \\ \hline
\end{tabular}
\end{table}

The initialization concerns the calculation of the background (according to Eqs.
(\ref{AL1})), the initial guess for the disturbed object $\mathbf{\tilde{u}}_{0}^{0}=\mathbf{\tilde{u}}_{0}^{init}$ (say, again by $AL$ \cite{MigukinAL}) and
Lagrange multipliers (e.g. $\mathbf{\Lambda }_{r}^{0}[k]=0$). Note that the
transform matrices for both the synthesis $\mathbf{\Psi }_{a}$,$\mathbf{\Psi 
}_{\varphi }$\ and analysis $\mathbf{\Phi }_{a}$,$\mathbf{\Phi }_{\varphi }$
may be constructed only once during the initialization procedure or may be
periodically updated. In this work we calculate these matrices only ones for
the compensated object amplitude $abs(\mathbf{\tilde{u}}_{0}^{0}[k])/abs(%
\mathbf{\hat{u}}_{B}[k])$ and phase estimates $angle(\mathbf{\tilde{u}}%
_{0}^{0}[k])-angle(\mathbf{\hat{u}}_{B}[k])$. Note also that in our experiments we use the $\ell _{1}$-norm in the sparse object approximation (``soft'' thresholding of the used BM3D filter, see \cite{MigukinSPIE12, Aram12}).

Note that the output of the $SPAR-BC$ phase-retrieval algorithm is not the
estimate of the disturbed $\mathbf{\tilde{u}}_{0}$ (Step 7), but the
estimate of the true object wave field $\mathbf{u}_{0}$\ calculated in Step
1. The derivations of main steps of $SPAR-BC$ (the minimization in Eqs. (\ref%
{opt2a1}),(\ref{opt2a2}),(\ref{opt1}) and (\ref{opt4})) can be found in our
previous works \cite{MigukinAL, VK12-2, MigukinSPIE12}. Step 4 returns the
wave field $\mathbf{u}_{r}^{t+1/2}$ at the sensor plane corresponding to the
forward propagation model with the optical mask $\mathbf{M}_{r}$. Step 5
gives the updates of $\mathbf{u}_{r}^{t+1/2}$ by their fitting to the
observations $\mathbf{o}_{r}$. The operator defining this update is denoted
as $\mathcal{G}$ and described in \cite[Appendix A]{MigukinAL}.

It is shown in \cite{VK12-2} that the object reconstruction with BM3D filtering
can be realized without Lagrange multipliers. However, it is found (see \cite{MigukinSPIE12}) that $\{\mathbf{\Lambda }_{r}\}$ help to recover small details of the object. In
Step 6 $\{\mathbf{\Lambda }_{r}^{t}\}$ are updated with the step $\alpha
_{r} $, and in our experiments we take a fixed step $\alpha _{r}=\alpha
=1/20 $ for all $K$ observations.

In this work we use the same noise variation at all sensor planes ($\sigma
_{r}^{2}=\sigma ^{2}$) and take the equal parameters for the Lagrangian
multipliers, $\gamma _{r}=\gamma $. Then, it is easy to see that the
estimate $\mathbf{\tilde{u}}_{0}^{t}$\ computed in Step 7 of $SPAR-BC$
consists of two parts: the disturbed object estimate calculated from the
observations and the filtered object approximation found from the output of
the BM3D filter. Then, this step of the algorithm can be given in the form%
\begin{equation}
\mathbf{\tilde{u}}_{0}^{t+1}=\sum_{r=1}^{K}\mathbf{B}_{r}\cdot (\mathbf{u}%
_{r}^{t+1}+\mathbf{\Lambda }_{r}^{t})+\kappa \cdot \mathbf{\tilde{v}}%
_{0}^{t+1},  \label{Eq13}
\end{equation}%
where $\mathbf{\tilde{v}}_{0}^{t+1}=\mathbf{\hat{u}}_{B}\circ \mathbf{v}%
_{0}^{t+1}$ and the transform matrix $\mathbf{B}_{r}$ is given in the form%
\begin{equation}
\mathbf{B}_{r}=(\sum_{r=1}^{K}\mathbf{A}_{r}^{H}\mathbf{A}_{r}+\kappa \cdot 
\mathbf{I}_{n\times n})^{-1}\mathbf{A}_{r}^{H},  \label{Eq13a}
\end{equation}%
and $\kappa =\gamma \sigma ^{2}/\gamma _{0}$. In particular, for all our
experiments $\kappa $ is equal to $90/25$.

\section{Numerical results}

\label{Sec5}\noindent In this Section a high performance of the proposed
algorithm is demonstrated by example of amplitude reconstructions from real
experimental data.

\subsection{Binary object model}

Here we consider reconstruction of a binary object with the amplitude given
as 
\begin{equation}
\mathbf{a}_{0}[k]=abs(\mathbf{u}_{0}[k])=\left\{ 
\begin{array}{c}
\beta _{1}\text{, for }k\in X_{1}\subset X\text{,} \\ 
\beta _{0}\text{, for }k\in X\backslash X_{1}\text{,}%
\end{array}%
\right.  \label{obj}
\end{equation}%
where $X$\ is a support of the image, $\beta _{0}\in 
\mathbb{R}
_{+}$\ and $\beta _{1}\in 
\mathbb{R}
_{+}$\ stand for the lower and upper level of the object amplitude signal,
respectively. The set $X_{1}$\ defines the indices of the upper level and
the set $X_{0}=X\backslash X_{1}$\ defines the indices of the lower level.
Both the levels $\beta _{0}$, $\beta _{1}$\ and the sets $X_{0}$, $X_{1}$\
are unknown and should be reconstructed. The U.S. Air Force resolution
test-chart is used for $\mathbf{a}_{0}$. For the amplitude-only object the
phase should be equal to zero, $angle(\mathbf{u}_{0}[k])=0$. In practice,
the laser beam passing through the chart undergoes some phase
transformations. These transformations define the phase characteristics of
the object $\mathbf{u}_{0}$, which are unknown. The only thing which can be
stated is that the pattern of the object phase reflects the binary features
of the amplitude model (\ref{obj}). Thus, we are looking for the
complex-valued object $\mathbf{u}_{0}$.

\subsection{Settings of parameters}

The standard settings of the phase-retrieval problem assume a multi-plane
lensless system with varying distances between the parallel object and
sensor planes. The intensity measurements at the sensor planes are used for
reconstruction of a 3D wave field including both the phase and amplitude.
Following \cite{Falldorf10} the considered 4f optical system works as an
imitator of this multi-plane lensless scenario. The principal difference is
that the sensor plane is immobile and fixed at the distance $4f$ from the
object plane. The effect of the varying distances is obtained by a phase
modulating SLM located at the Fourier plane, where different optical masks $%
\mathbf{M}_{r}$\ corresponding to the propagation distances are programmed
as 
\begin{eqnarray}
&&\mathbf{M}_{r}[\eta _{1},\eta _{2}]=  \label{DTF} \\
&=&\exp (2i\frac{\pi }{\lambda }z_{r}\sqrt{1-\Delta _{v_{1}}^{2}\frac{|\eta
_{1}|_{2}^{2}}{f^{2}}-\Delta _{v_{2}}^{2}\frac{|\eta _{2}|_{2}^{2}}{f^{2}}})
\notag
\end{eqnarray}

In Eq. (\ref{DTF}) $z_{r}=z_{1}+(r-1)\cdot \Delta _{z},$ $r=1,...K$ are the
distances between the object and sensor planes. In our experiments $K$=5, $%
z_{1}$=20$mm$ is the distance from the object to the first measurement
plane, $\Delta _{z}$=2$mm$ is the fixed distance between successive
measurement planes.

Due to the bandlimitedness (fixed size of the SLM) and discrete
representation of the optical mask (on a 2D array of liquid crystal cells of
the SLM), the experimental results at the sensor plane will be different
from the output of the model calculated using the angular spectrum
decomposition (ASD, \cite{Goodman}). These differences are consider as
components of the background to be estimated and compensated.

In our discrete wave field propagation model, the pixels at the sensor and
Fourier planes are square of the different size $\Delta _{y_{1}}\times
\Delta _{y_{1}}$=3.45$\times $3.45 ($\mu m$) and $\Delta _{v_{1}}\times
\Delta _{v_{2}}$=8$\times $8 ($\mu m$), respectively, with 100\% fill factor 
\cite{Arizon, agour10}. The object is pixelated with the sensor size pixels: 
$\Delta _{x_{1}}=\Delta _{x_{2}}=\Delta _{y_{1}}=\Delta _{y_{2}}$. The
transparent U.S. Air Force resolution test-chart (MIL-STD-150A) inserted in
the front focal plane of the first lens $L_{1}$ is illuminated by collimated
coherent light with wavelength $\lambda $=532 $nm$ (i.e. a green Nd:YAG
laser is used). The employed SLM was supplied by Holoeye Photonics AG and
configured to provide full $2\pi $ phase modulation. The focal distance of
lenses used in the 4f configuration is $f$=150 $mm$ what equates with the
image size $2892\times 2892$ pixels according to the sampling conditions, Eq. (\ref{sc1}). The measurement area is smaller and here we reconstruct only a
part of the object of the size $N_{x}\times N_{y}$ ($2048\times 2048$) for
the corresponding computational focal distance $f_{c}$=106.25 $mm$ (see Eqs. (\ref{F01})--(\ref{F02})). Note that $f$=150 $mm$ in Eq. (\ref{DTF}) defining
the optical masks $\mathbf{M}_{r}$.

The algorithm is implemented for a graphic processing unit (GPU) in order to
use the advantage of parallel processing of $\mathbf{u}_{r}^{t+1}$\ and $%
\mathbf{u}_{0}^{t+1}$. The GPU realization results in a significant
acceleration what is crucial especially for large images \cite{MigukinSPIE12}%
. The presented results are computed in MATLAB 7.13 (R2011b) using GPU
Nvidia GF460GTX with CUDA 4.1. The computer used for experiments is Intel i5
2500 (4 physical cores) at 3.3 GHz; 8Gb RAM, Windows 7 SP1.

\subsection{Modification of $SPAR-BC$ for binary object}

A special modification of the filtering procedure is developed targeted to
improve reconstruction of a binary object. The BM3D filtering (Step 2 of $%
SPAR-BC$) is replaced by%
\begin{eqnarray}
\mathbf{a}_{0}^{t+1/3} &=&BM3D_{a}(abs(\mathbf{u}_{0}^{t})-\beta
_{0}^{t})+\beta _{0}^{t}  \label{aa} \\
\mathbf{a}_{0}^{t+1/2} &=&BM3D_{a}(\mathbf{a}_{0}^{t+1/3}-\beta
_{1}^{t})+\beta _{1}^{t}  \notag
\end{eqnarray}%
where $\beta _{0}^{t}$ and $\beta _{1}^{t}$ are scalar variables. These
variables are calculated as medians of $\mathbf{a}_{0}^{t}=abs(\mathbf{u}%
_{0}^{t})$ over the sets: $X_{0}^{t}=\{\mathbf{a}_{0}^{t}:0\leq \mathbf{a}%
_{0}^{t}\leq \rho ^{t}\}$ and $X_{1}^{t}=\{\mathbf{a}_{0}^{t}:\mathbf{a}%
_{0}^{t}>\rho ^{t}\}$ 
\begin{eqnarray}
\beta _{0}^{t} &=&median_{\mathbf{a}_{0}^{t}\in X_{0}^{t}}(\mathbf{a}%
_{0}^{t}),  \label{aa2} \\
\beta _{1}^{t} &=&median_{\mathbf{a}_{0}^{t}\in X_{1}^{t}}(\mathbf{a}%
_{0}^{t})  \notag
\end{eqnarray}

In order to estimate these classes $X_{0}^{t}$ and $X_{1}^{t}$, corresponding to
small and large values of $\mathbf{a}_{0}^{t}$, we use the thresholding
parameter $\rho ^{t}$ calculated using the Otsu algorithm \cite{Otsu}.\ In
the procedures (\ref{aa}), successive subtractions of $\beta _{0}^{t}$ and $%
\beta _{1}^{t}$ makes the image flatter first in the area of low values of
binary amplitude signal and after that in the area of its high values.
Experiments show that this flattening enables much more efficient filtering
of artifacts for the estimate of $\mathbf{a}_{0}$ in case of binary object.

For the phase filtering we make the flattering procedure simpler because for
the considered $\mathbf{u}_{0}$\ the phase should be close to zero. The
median of the object phase is calculated only ones as $\mathbf{\varphi }%
_{m}^{t}=median(angle(\mathbf{u}_{0}^{t}))$ without partitioning in two
subsets as for the object amplitude. Finally, the filtering procedure by the
BM3D filter (Steps 2 and 3 in $SPAR-BC$) is replaced by the following ones%
\begin{eqnarray}
\mathbf{a}_{0}^{t+1/3} &=&BM3D_{a}(abs(\mathbf{u}_{0}^{t})-\beta
_{0}^{t})+\beta _{0}^{t}  \label{aa3} \\
\mathbf{a}_{0}^{t+1/2} &=&BM3D_{a}(\mathbf{a}_{0}^{t+1/3}-\beta
_{1}^{t})+\beta _{1}^{t}  \notag \\
\mathbf{\varphi }_{0}^{t+1/2} &=&BM3D_{\varphi }(angle(\mathbf{u}_{0}^{t})-%
\mathbf{\varphi }_{m}^{t})  \notag \\
\mathbf{v}_{0}^{t+1} &=&\mathbf{a}_{0}^{t+1/2}\circ \exp (j\cdot (\mathbf{%
\varphi }_{0}^{t+1/2}+\mathbf{\varphi }_{m}^{t}))  \notag
\end{eqnarray}

The presented results of the object reconstruction from experimental data
are obtained using this modification BM3D filtering.

\subsection{Initialization for $SPAR-BC$: reconstruction without background compensation}

\begin{figure}[t!]
\begin{center}
\includegraphics[scale=0.12]{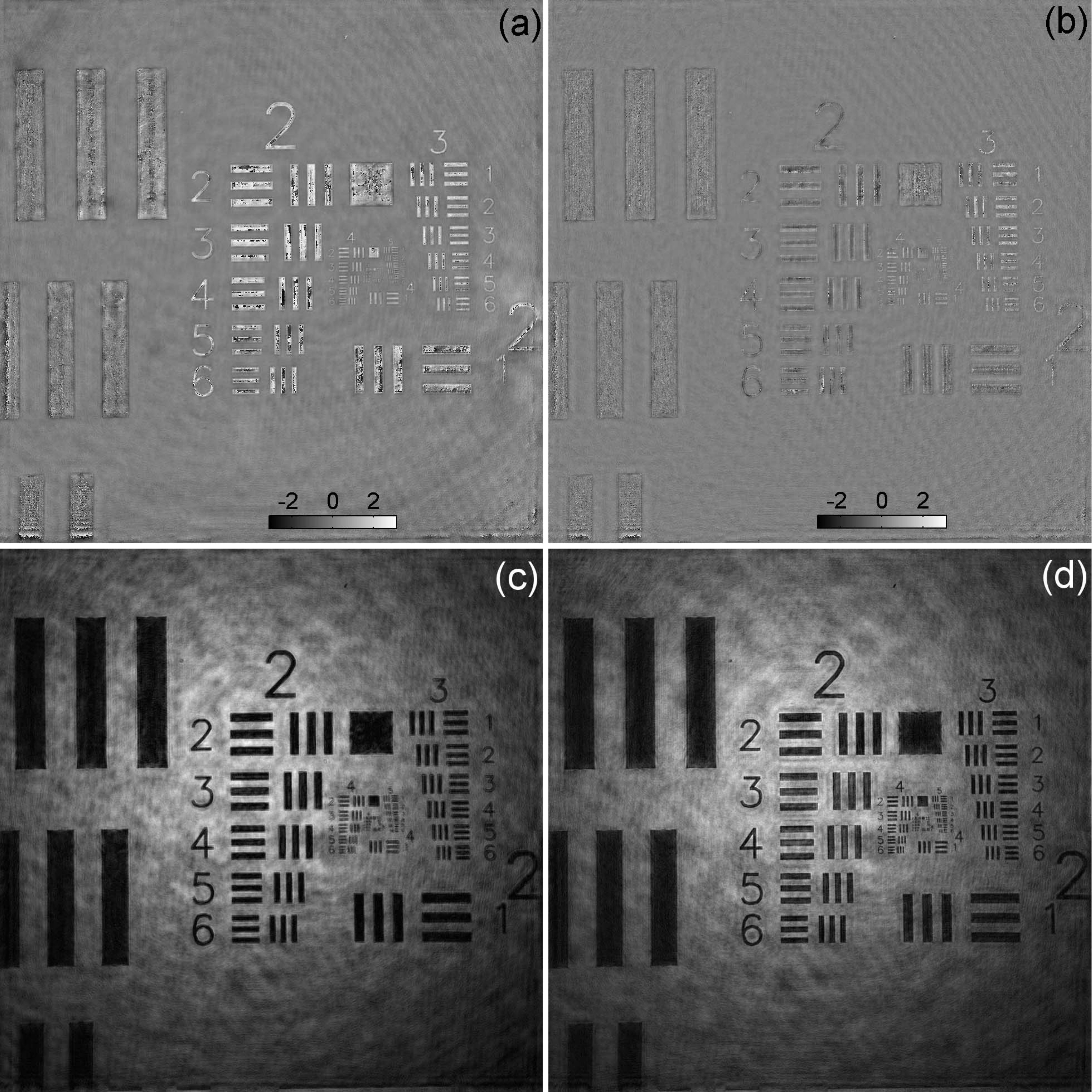}
\end{center}
\caption{ Reconstructions of the ``disturbed'' object computed from experimental data (left column) by $AL$ and (right column) by FA\ \protect\cite{AgourSPIE}. In the top row the amplitude reconstructions are presented: (a) by $AL$, (c) by FA. In the bottom row we demonstrate the phase estimates: (c) by $AL$, (d) by FA. The object reconstruction by $AL$ ($\mathbf{\tilde{a}}^0_0$ from (a) and $\mathbf{\tilde{\protect\varphi}}^0_0$ from (c)) is used for the initialization $\mathbf{\tilde{u}}_{0}^{0} = \mathbf{\tilde{a}}^0_0 \circ \func{exp}(j \cdot \mathbf{\tilde{%
\protect\varphi}}^0_0)$ in the $SPAR-BC$ algorithm. }
\label{Fig05}
\end{figure}

Here we consider the object reconstruction from the experimental data
obtained by the approach originated in \cite{Falldorf10, Agour}. The
complex amplitude of the disturbed object is obtained by two various
algorithm: by the mentioned $AL$ algorithm \cite{MigukinAL} and by the
successive phase-retrieval algorithm described in \cite{AgourSPIE}. The
second iterative algorithm is close to the circular wave reconstruction
originated in \cite{Pedrini, Almoro-Pedrini}: the calculated amplitude is
replaced by the square root of the given noisy intensity, keeping the
calculated phase (the initial guess for the phase is zero). For simplicity,
we refer to the latter algorithm as the Falldorf-Agour (FA) algorithm.

In Fig. \ref{Fig05} we present the comparison of the reconstruction imaging
of the disturbed object obtained from experimental data by these different
methods\textbf{.} In the right column the estimate of the disturbed object
phase and amplitude computed by the FA algorithm \cite{AgourSPIE}\ are
illustrated (see Fig. \ref{Fig05}(b) and Fig. \ref{Fig05}(d), respectively).
In the left column the reconstructed disturbed object phase and amplitude found by $AL$ \cite%
{MigukinAL}\ are shown (see Fig. \ref{Fig05}(a) and Fig. \ref{Fig05}(c),
respectively). These results are shown for 25 iterations of the
phase-retrieval algorithms. The artifacts which definitely should be
addressed to the background are clearly seen in these images. It can be seen
that the amplitude estimate by $AL$\ is significantly oversmoothed comparing
with the result by FA. It is manifested in partial suppression of the
diffraction artifacts on the geometrical elements with some degradation of a
smooth surface as well. The phase reconstructions are not flat and have
certain errors in the regions of the digits and geometrical figures in the
amplitude. The phase by $AL$ has stronger degradation of the phase comparing
with the phase calculated by \cite{AgourSPIE}, because of a weak correction
of the object reconstruction by Lagrange multipliers. We take quite small $%
\alpha =1/20$ because we are looking for a sharp result. It is found that
larger $\alpha $\ may correct the phase estimate but leads to more noisy
amplitude reconstruction, and denoising by larger regularization parameter $%
\mu $\ (see Eq. (\ref{AL0})) results in oversmoothing.

\begin{figure}[t!]
\begin{center}
\includegraphics[scale=0.12]{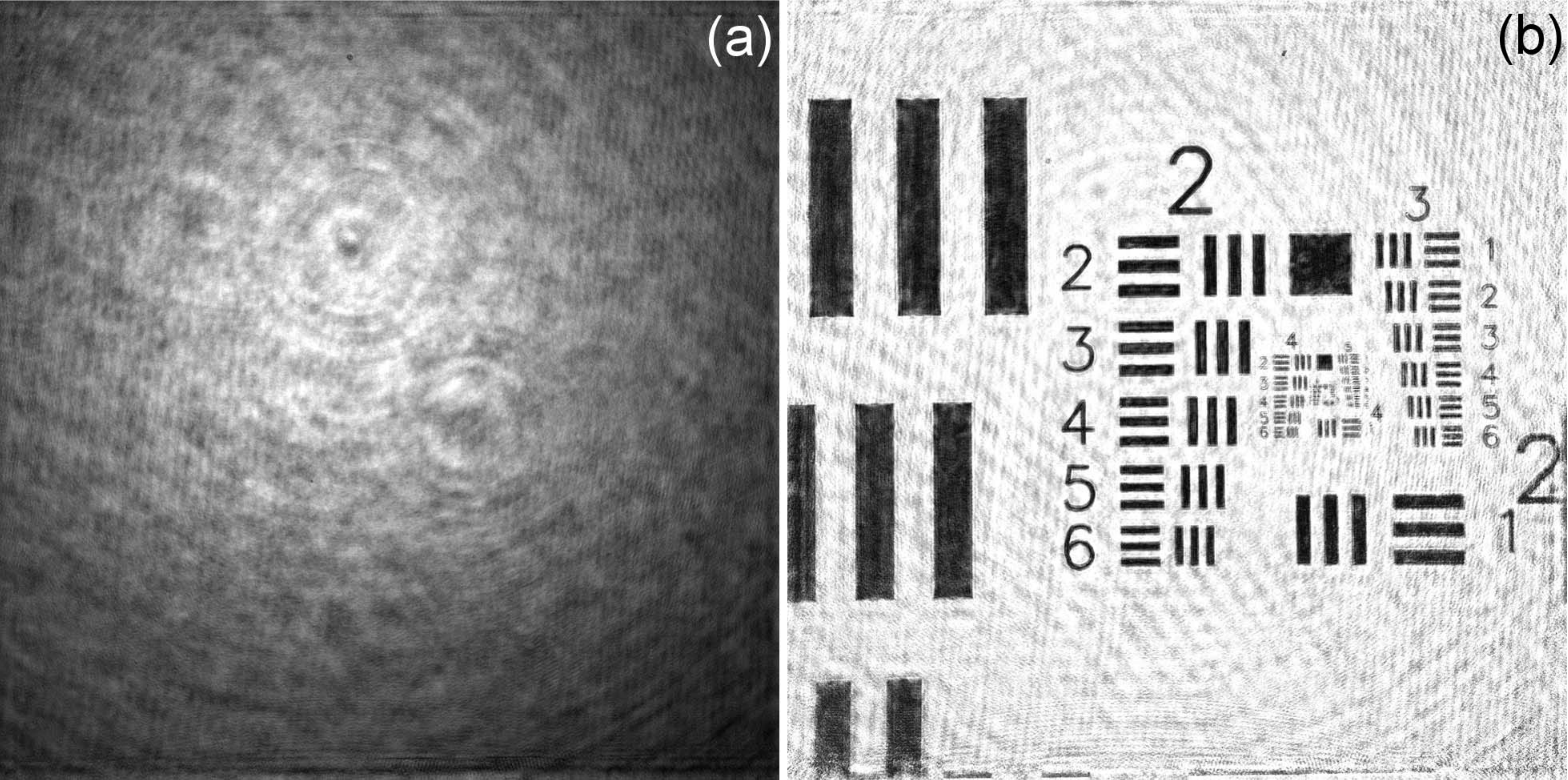}
\end{center}
\caption{ Background compensation in $SPAR-BC$: (a) the reconstructed amplitude of the background $abs(\mathbf{\hat{u}}_{B})$ and (b) the initial guess for the object amplitude \  $abs(\mathbf{u}_{0}^{0})$ found with the smoothed background amplitude $BM3D_{a}(abs(\mathbf{\hat{u}}_{B}))$. }
\label{Fig06}
\end{figure}

\begin{figure}[b!]
\begin{center}
\includegraphics[scale=0.32]{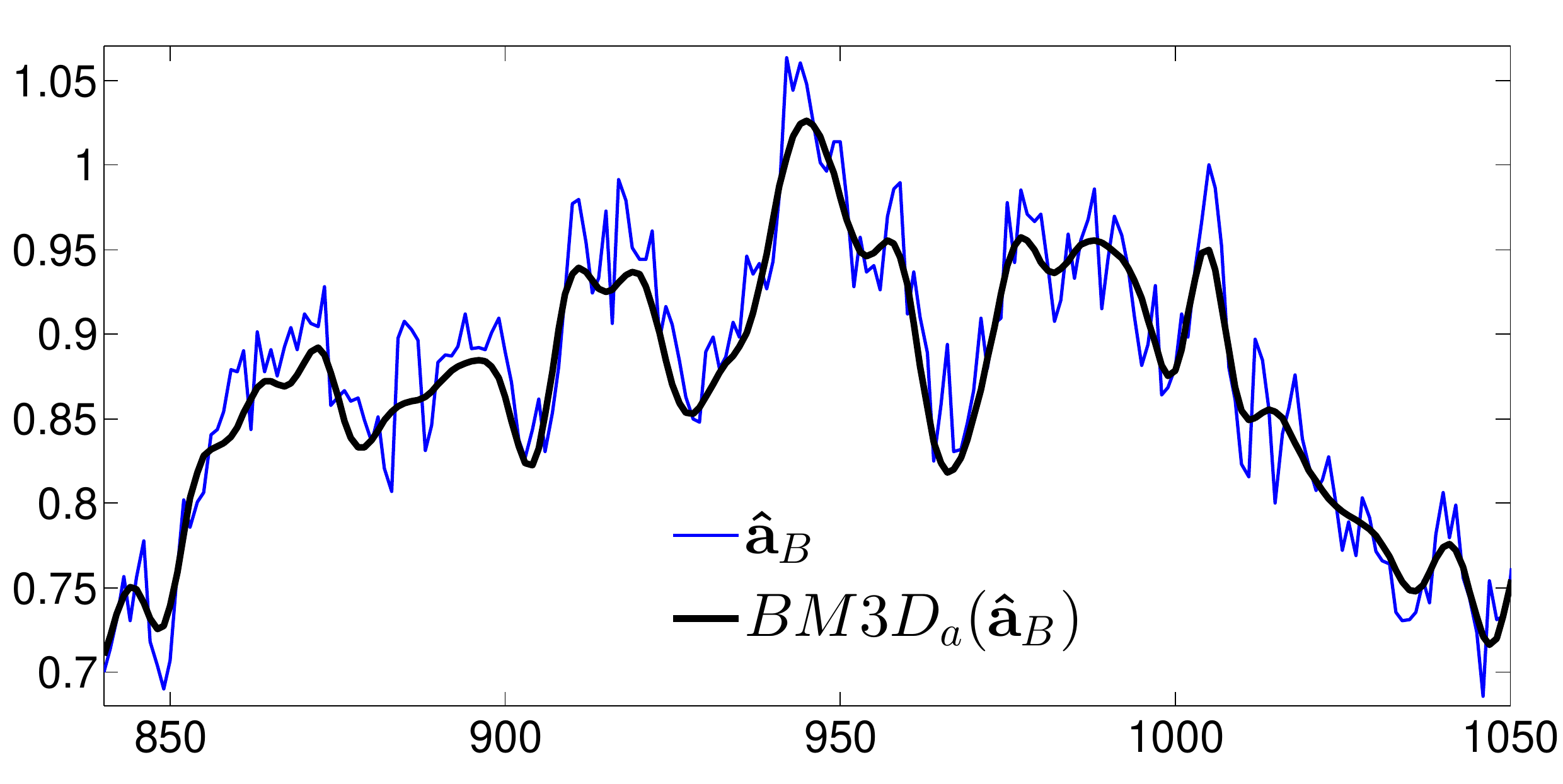}
\end{center}
\caption{ Cross-sections of (thin curve) the amplitude estimate of the background, $\mathbf{\hat{a}}_{B}=abs(\mathbf{\hat{u}}_{B})$, and (thick curve)\ its smoothed version computed by BM3D filter, $BM3D_{a}(\mathbf{\hat{a}}_{B})$. }
\label{Fig07}
\end{figure}

Note that the imperfect $AL$ object estimate is used only as an
initialization $\mathbf{\tilde{u}}_{0}^{0} = \mathbf{\tilde{a}}^0_0 \circ 
\func{exp}(j \cdot \mathbf{\tilde{\varphi}}^0_0)$ for the main procedure of $%
SPAR-BC$.

\subsection{$SPAR-BC$: reconstruction with background compensation}

The background reconstruction is produced with the calibration experiments
for the free space object $\mathbf{u}_{0}[k]=1$. Figure \ref{Fig06}(a)
demonstrates the reconstructed background amplitude. However, the obtained
results are appeared quite noisy and an additional postfiltering of the
background amplitude $\mathbf{\hat{a}}_{B}$ is introduced. Here we use the
smoothed version of this background reconstruction with no high frequency
components. For the filtering of the amplitude and the phase of $\mathbf{%
\hat{u}}_{B}$ we again use BM3D filter \cite{Aram}. The cross-sections of
the original $\mathbf{\hat{a}}_{B}=abs(\mathbf{\hat{u}}_{B})$ and smoothed $%
\mathbf{\tilde{a}}_{B}=BM3D_{a}(\mathbf{\hat{a}}_{B})$ background amplitudes
are illustrated in Fig. \ref{Fig07}. The result of the compensation of the
initial object amplitude by such an smoothed background $\mathbf{a}%
_{0}^{0}[k]=abs(\mathbf{\tilde{u}}_{0}^{0}[k])/\mathbf{\tilde{a}}_{B}[k]$ is
shown in Fig. \ref{Fig06}(b). The corresponding cross-section is presented in Fig. \ref{Fig09}.

\begin{figure}[t!]
\begin{center}
\includegraphics[scale=0.16]{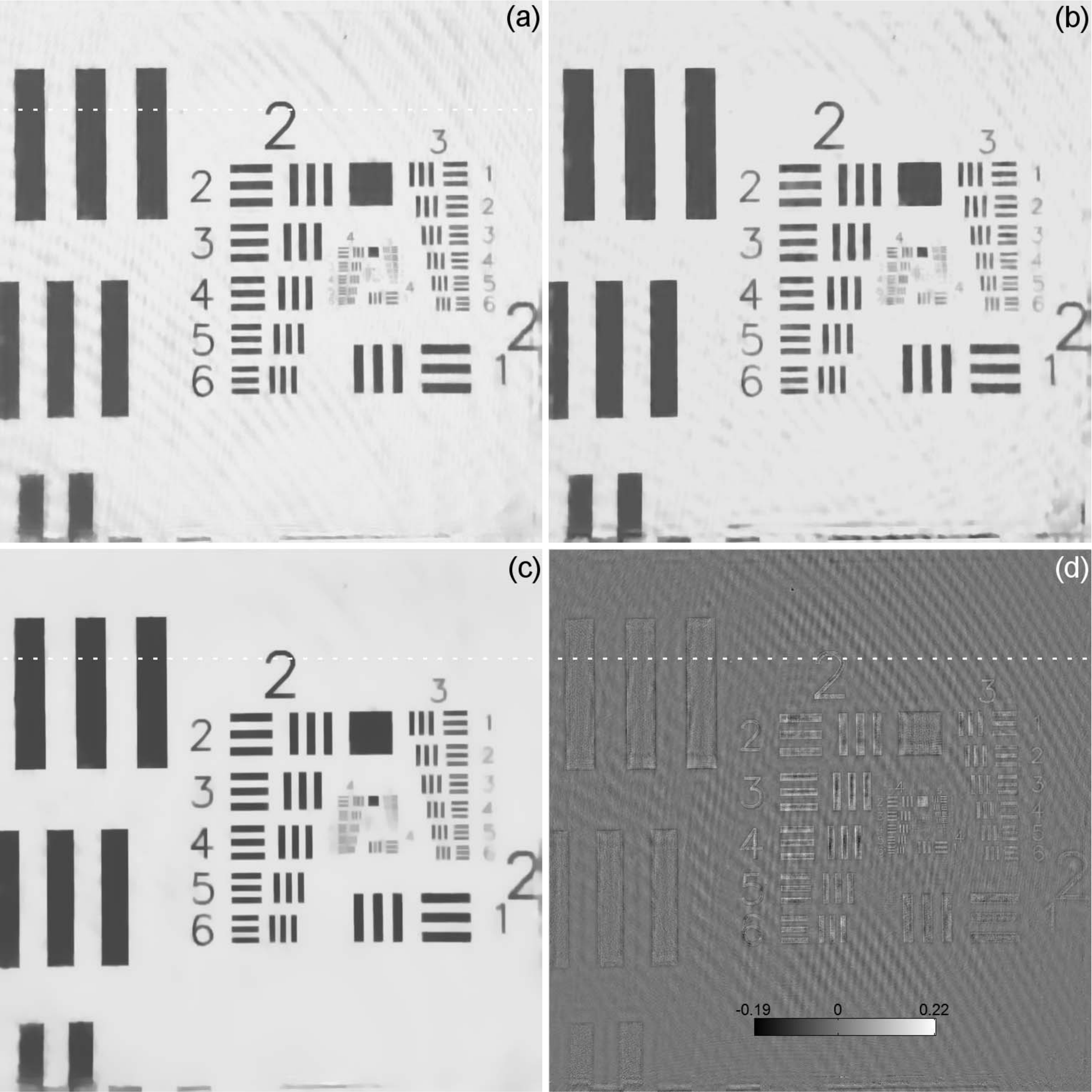}
\end{center}
\caption{ Object reconstruction by $SPAR-BC$. Comparison of the reconstructed object amplitudes $abs(\mathbf{\hat{u}}_{0})$ calculated (a) with the smoothed background estimate and (b) with the original background reconstruction. The result of postfiltering of the object amplitude $BM3D_{a}(abs(\mathbf{\hat{u}}_{0}))$, $\protect\tau _{a}\protect\gamma _{a}=0.04$, is demonstrated in (c). The object phase estimate\ $angle(\mathbf{\hat{u}}_{0})$ is illustrated in (d). The noise in the filtered phase estimate is totally suppressed. }
\label{Fig08}
\end{figure}

\begin{figure}[b!]
\begin{center}
\includegraphics[scale=0.32]{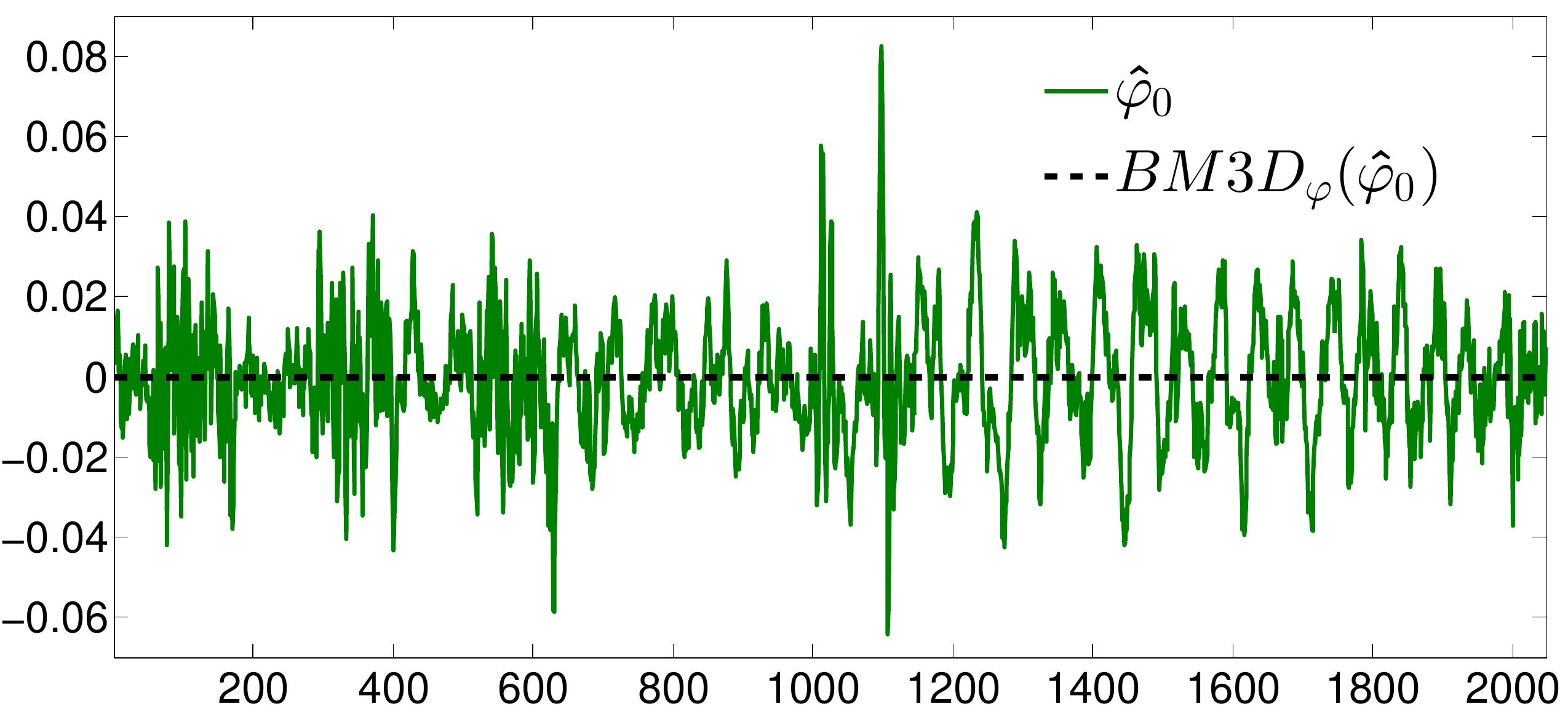}
\end{center}
\caption{ Cross-sections of the reconstructed object phase $\mathbf{\hat{\protect\varphi}}_{0}=angle(\mathbf{\hat{u}}_{0})$ and the filtered phase $BM3D_{\protect\varphi }(\mathbf{\hat{\protect\varphi}}_{0})$, $\protect\tau _{\protect\varphi }\protect\gamma _{\protect\varphi }=0.08$. These results are shown along the dashed lines shown in Fig. \protect\ref{Fig08}(d).}
\label{Fig10}
\end{figure}

The results of the wave field reconstruction obtained by the developed $%
SPAR-BC$ algorithm are shown in Fig. \ref{Fig08} for 25 iterations. In Figs. %
\ref{Fig08}(a) and \ref{Fig08}(b) the object amplitude reconstruction
calculated with the smoothed and original background estimates are presented, respectively. It can be seen that the imaging
of the amplitude estimate obtained using the smoothed background (Fig. \ref%
{Fig08}(a)) is essentially better comparing with the reconstruction found
with the background without postfiltering, namely: the artifacts are well
seen of the border of Fig. \ref{Fig08}(b). Further improvement of the imaging can be achieved by postfiltering of the reconstructed $\mathbf{\hat{u}}_{0}$, and in Fig. \ref{Fig08}(c) the result of such an additional BM3D\ filtering of $\mathbf{\hat{u}}_{0}$ (computed again with the smoothed $\mathbf{\tilde{a}}_{B}$) is demonstrated. The reconstructed object phase $angle(\mathbf{\hat{u}}_{0})$
is illustrated in Fig. \ref{Fig08}(d).

The threshold parameter of the BM3D filtering in $SPAR-BC$ is $\tau_{a}\gamma _{a}=0.13$ for the object amplitude and $\tau _{\varphi }\gamma
_{\varphi }=2$\ for the phase. Note that even with a large thresholding $\tau _{\varphi }\gamma _{\varphi }=2$ we have significant noise in the phase estimate (see Fig. \ref{Fig08}(d)). Taking into account that $\mathbf{\varphi }_{0}[k]=0$, the
reconstruction accuracy for the phase is RMSE=0.2. However, we can completely wipe the phase noise out by the mentioned additional filtering by BM3D with quite a
small $\tau _{\varphi }\gamma _{\varphi }=0.08$: compare the cross-sections
of $\mathbf{\hat{\varphi}}_{0}$ and $BM3D_{\varphi }(\mathbf{\hat{\varphi}}%
_{0})$ in Fig. \ref{Fig10}.

It is shown in Eq. (\ref{Eq13}) that the object reconstruction is a weighted
sum of a noisy estimate from the propagation model and filtered
approximation from the previous iteration. Thus, at each iteration the
filtered object estimate is corrupted due to the used noisy intensity
observations. It is a challenge to find a proper balance between denoising
and oversmoothing. We obtain quite a sharp results of $\mathbf{\hat{a}}_{0}$
with some small distortions (see Fig.\ref{Fig08}(a) and the related
cross-section in Fig. \ref{Fig09}). However remaining noise and artifacts
can be additionally suppressed by BM3D filter after the main procedure of $%
SPAR-BC$. It provides crisp imaging, but the resulting amplitude is
oversmoothed and small details are almost lost: the result of postfiltering
of $\mathbf{\hat{a}}_{0}$ by 12 iterations with $\tau _{a}\gamma _{a}=0.04$
is presented in Fig. \ref{Fig08}(c) and the corresponding cross-section in
Fig. \ref{Fig09}.

\begin{figure}[t!]
\begin{center}
\includegraphics[scale=0.35]{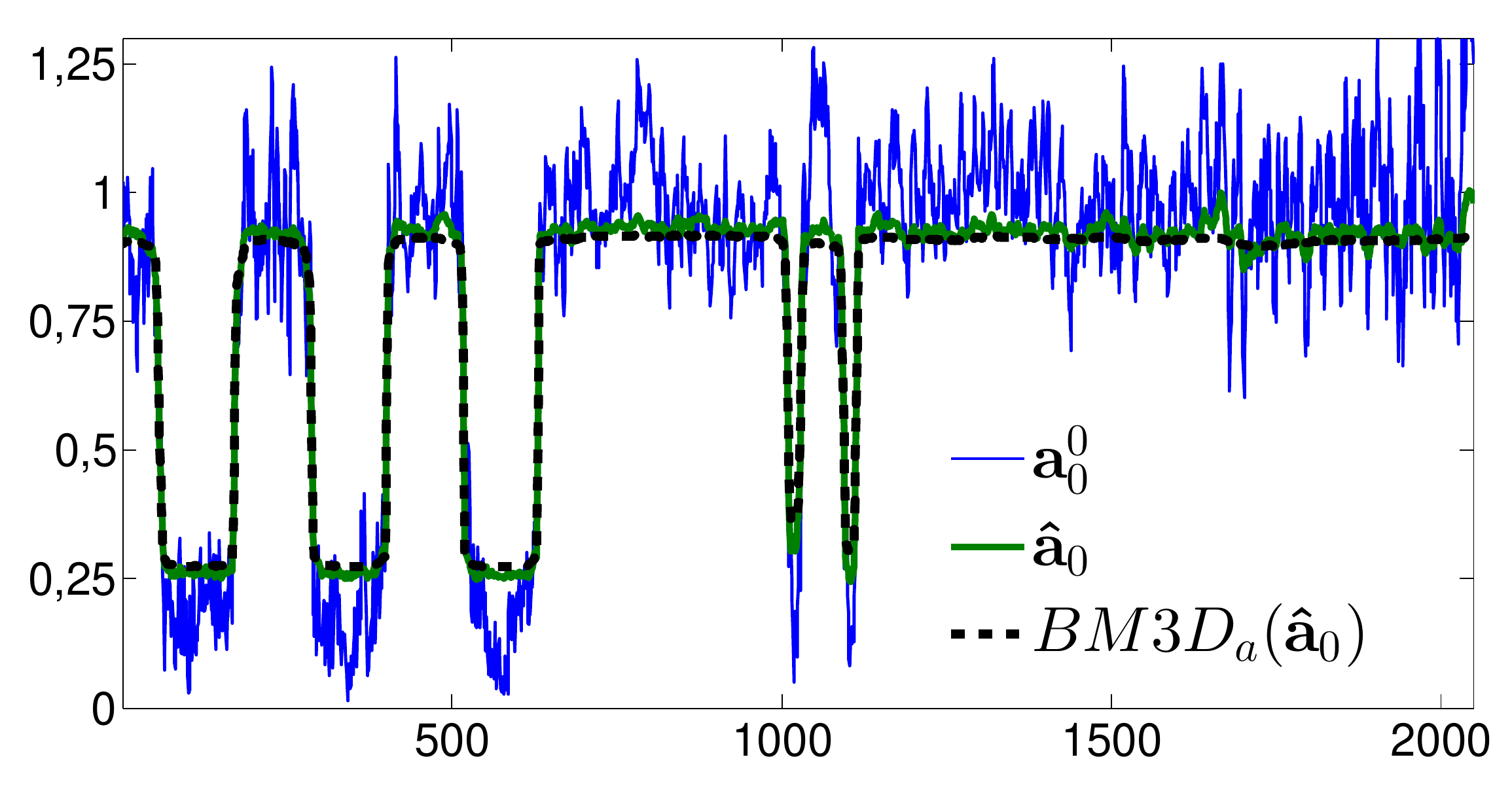}
\end{center}
\caption{ Cross-sections of (solid thin curve) the initial guess $\mathbf{a}%
_{0}^{0}=abs(\mathbf{u}_{0}^{0})$, (solid thick) the reconstructed amplitude 
$\mathbf{\hat{a}}_{0}=abs(\mathbf{\hat{u}}_{0})$\ and (dashed curve) the
filtered amplitude estimate $BM3D_{a}(\mathbf{\hat{a}}_{0})$. These results
are related to the imaging presented in Figs. \protect\ref{Fig06}(b), 
\protect\ref{Fig08}(a) and \protect\ref{Fig08}(c), respectively. The
cross-sections are given along the dashed line in Fig. \protect\ref{Fig08}.}
\label{Fig09}
\end{figure}

\section{Discussion and conclusion}

It can be seen that the modified BM3D filtering (Eqs. (\ref{aa3}))\ works
here as a classifier for the noisy binary object estimate. The estimate of
the amplitude levels are found using the Otsu method, but the BM3D filtering
shifts the value of the pixel $\mathbf{a}_{0}^{t}[k]$ to one of these two
levels $\beta _{0}^{t}$ or $\beta _{1}^{t}$ depending on the local
neighborhood.

\begin{figure}[t!]
\begin{center}
\includegraphics[scale=0.35]{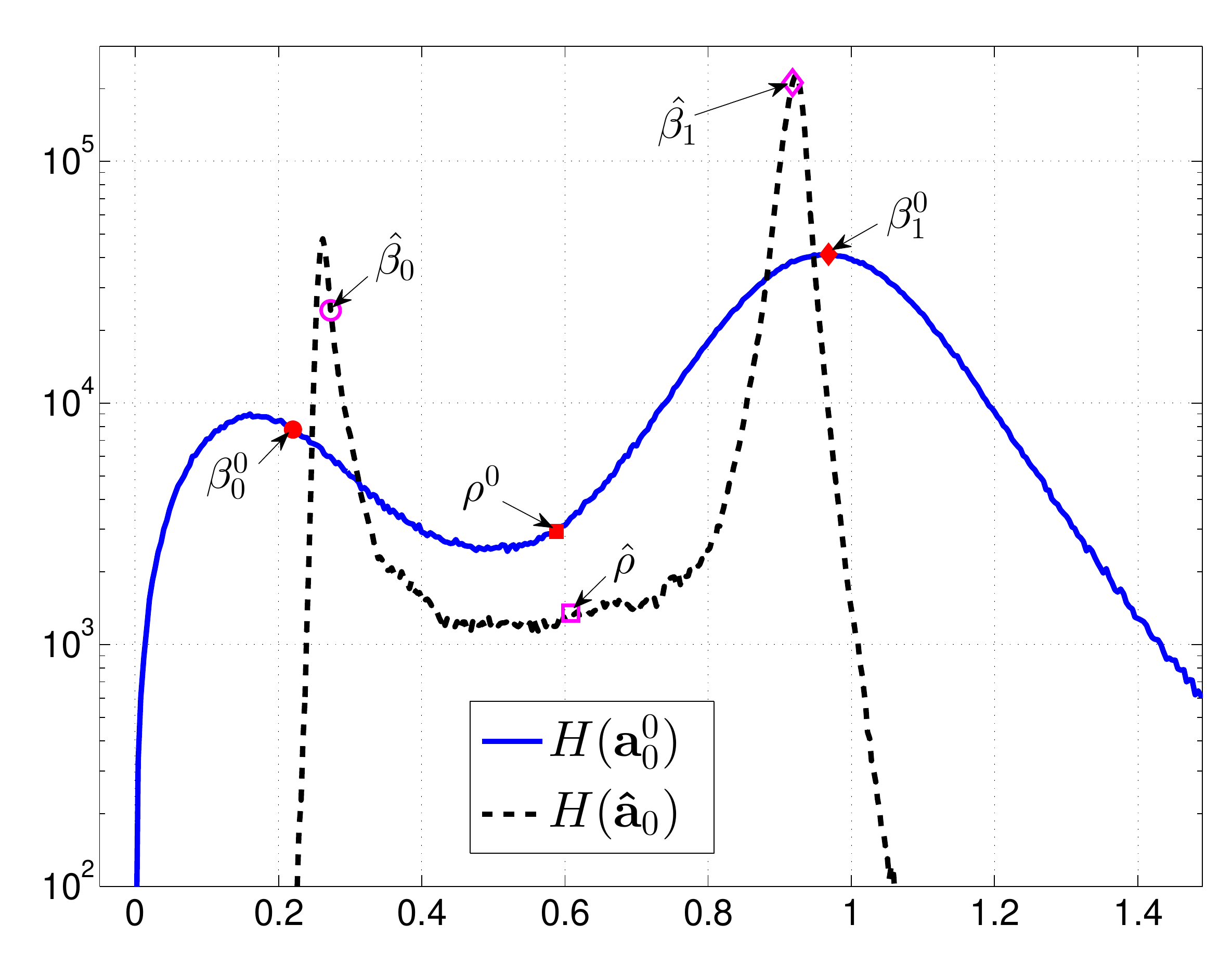}
\end{center}
\caption{ Partition produced according to the modified BM3D filtering (Eqs. (\protect\ref{aa3})). (solid curve) $H(\mathbf{a}_{0}^{0})$ is the histogram for the initial estimate of the object amplitude, $\mathbf{a}_{0}^{0}=abs(\mathbf{u}_{0}^{0})$\ presented in Fig. \protect\ref{Fig06}(b). (dashed curve) $H(\mathbf{\hat{a}}_{0})$ is the histogram for the resulting object amplitude estimate after 25 iterations, $\mathbf{\hat{a}}_{0}=abs(\mathbf{\hat{u}}_{0})$ illustrated in Fig. \protect\ref{Fig08}(a). Otsu's threshold with the upper and lower levels of the initial ($\protect\rho ^{0},\protect\beta _{1}^{0},\protect\beta _{0}^{0}$) and resulting object estimates ($\hat{\protect\rho},\hat{\protect\beta}_{1},\hat{\protect\beta}_{0}$) are also presented. }
\label{FigX1}
\end{figure}

The result of classification is presented in Fig. \ref{FigX1}. Let $H(\cdot)$ stands for a histogram of a discrete distribution.
The histogram $H(\mathbf{a}_{0}^{0})$\ for the initial estimate of the
compensated object amplitude is denoted in Fig. \ref{FigX1} by a solid
curve. It is already seen that the object $\mathbf{a}_{0}$ looks to be
binary. The histogram of the resulting $\mathbf{\hat{a}}_{0}$ is denoted
here by a dashed curve. In Fig. \ref{FigX1}\ we also present the initial guess
of the lower $\beta _{0}^{0}$\ and upper levels $\beta _{1}^{0}$, the
resulting lower $\hat{\beta}_{0}$\ and upper levels $\hat{\beta}_{1}$, and
the Otsu threshold for the initialization $\rho ^{0}$\ and the final step $%
\hat{\rho}$ of our reconstruction. Note that blurred regions (see small
details and borders on the geometrical elements e.g. in Fig. \ref{Fig08}(a))
correspond to a \textquotedblleft sloping valley\textquotedblright\ between
two peaks of $H(\mathbf{\hat{a}}_{0})$ in Fig. \ref{FigX1}.

In this paper a novel phase-retrieval algorithm with background compensation and powerful BM3D filtering is presented. The $SPAR-BC$ algorithm demonstrates a very good reconstruction quality : we have a clear separation of the binary true object, and the background estimate \textquotedblleft undertakes\textquotedblright\ strong fluctuations, which would be difficult to compensate by filtering only. The reconstructions by two different phase-retrieval methods ($AL$ and FA) are presented to emphasize the obtained enhancement of imaging of the developed algorithm with respect to modern phase-retrieval algorithms with no background compensation (compare the results in Figs.\ \ref{Fig05} and Fig. \ref{Fig08}).


\begin{thebibliography}{99}
\small

\bibitem{GS72} R. W. Gerchberg and W. O. Saxton, \textquotedblleft A
practical algorithm for the determination of phase from image and
diffraction plane pictures,\textquotedblright\ \textit{Optik} \textbf{35}, 237--246
(1972).

\bibitem{Missel73} D. L. Misell, \textquotedblleft A method for the solution
of the phase problem in electron microscopy,\textquotedblright\ \textit{J. Phys. D}
\textbf{6}, L6--L9 (1973).

\bibitem{Gonsalves} R. A. Gonsalves, \textquotedblleft Phase retrieval from
modulus data,\textquotedblright\ \textit{J. Opt. Soc. Am.} \textbf{66}, 961--964
(1976).

\bibitem{Fienup78} J. R. Fienup, \textquotedblleft Reconstruction of an
object from the modulus of its Fourier transform,\textquotedblright\ \textit{Opt. Lett.} \textbf{3}, 27--29 (1978).

\bibitem{YG81} B. Gu and G. Yang, \textquotedblleft On the phase retrieval
problem in optical and electronic microscopy,\textquotedblright\ \textit{Acta Opt. Sin.} \textbf{1}, 517--522 (1981).

\bibitem{YG94} G. Yang, B. Dong, B. Gu, J. Zhuang, and O. K. Ersoy,
\textquotedblleft Gerchberg-Saxton and Yang-Gu algorithms for phase
retrieval in a nonunitary transform system: a comparison,\textquotedblright\ \textit{Appl. Opt.} \textbf{33}, 209--218 (1994).

\bibitem{Z96} Z. Zalevsky, D. Mendlovic, and R. Dorsch, \textquotedblleft
Gerchberg--Saxton algorithm applied in the fractional Fourier or the Fresnel
domain,\textquotedblright\ \textit{Opt. Lett.} \textbf{21}, 842--844 (1996).

\bibitem{Gureyev3} T. E. Gureyev, \textquotedblleft Composite techniques for
phase retrieval in the Fresnel region,\textquotedblright\ \textit{Opt. Commun.} 
\textbf{220}, 49--58 (2003).

\bibitem{Fienup80} J. R. Fienup, \textquotedblleft Iterative method applied
to image reconstruction and to computer generated
holograms,\textquotedblright\ \textit{Opt. Eng.} \textbf{19}, 297--305 (1980).

\bibitem{Fienup82} J. R. Fienup, \textquotedblleft Phase retrieval
algorithms: a comparison,\textquotedblright\ \textit{Appl. Opt.} \textbf{21},
2758--2769 (1982).

\bibitem{Ivanov} V. Yu. Ivanov, V. P. Sivokon, and M. A. Vorontsov,
\textquotedblleft Phase retrieval from a set of intensity measurements:
theory and experiment,\textquotedblright\ \textit{J. Opt. Soc. Am. A} \textbf{9},
1515--1524 (1992).

\bibitem{Pedrini} G. Pedrini, W. Osten, and Y. Zhang, \textquotedblleft
Wave-front reconstruction from a sequence of interferograms recorded at
different planes,\textquotedblright\ \textit{Opt. Lett.} \textbf{\ 30}, 833--835
(2005).

\bibitem{Almoro-Pedrini} P. Almoro, G. Pedrini, and W. Osten,
\textquotedblleft Complete wavefront reconstruction using sequential
intensity measurements of a volume speckle field,\textquotedblright\ \textit{Appl. Opt.} \textbf{45}, 8596--8605 (2006).

\bibitem{Goodman} J. W. Goodman, \textit{Introduction to Fourier Optics},
2nd ed. (McGraw-Hill, 1996).

\bibitem{Falldorf10} C. Falldorf, M. Agour, C. v. Kopylow, and R. B.
Bergmann, \textquotedblleft Phase retrieval by means of a spatial light
modulator in the Fourier domain of an imaging system,\textquotedblright\ 
\textit{Appl. Opt.} \textbf{49}, 1826--1830 (2010).

\bibitem{Xue} Q. Xue, Z. Wang, J. Huang, and J. Gao, \textquotedblleft The
elimination of the errors in the calibration image of 3D measurement with
Structured Light,\textquotedblright\ \textit{Proc. SPIE} \textbf{8430}, (2012).

\bibitem{Cuche2000} E. Cuche, P. Marquet, and C. Depeursinge,
\textquotedblleft Spatial filtering for zero-order and twin-image
elimination in digital off-axis holography,\textquotedblright\ \textit{Appl. Opt.} \textbf{39}, 4070--4075 (2000).

\bibitem{Ferraro03} P. Ferraro, S. D. Nicola, A. Finizio, G. Coppola, S.
Grilli, C. Magro, and G. Pierattini, \textquotedblleft Compensation of the
inherent wave front curvature in digital holographic coherent microscopy for
quantitative phase-contrast imaging,\textquotedblright\ \textit{Appl. Opt.} \textbf{42%
}, 1938--1946 (2003).

\bibitem{Pedrini01} G. Pedrini, S. Schedin, and H. J. Tiziani,
\textquotedblleft Aberration compensation in digital holographic
reconstruction of microscopic objects,\textquotedblright\ \textit{J. Mod. Opt.} \textbf{48}, 1035--1041 (2001).

\bibitem{Zhao12} S. M. Zhao, J. Leach, L. Y. Gong, J. Ding, and B. Y. Zheng,
\textquotedblleft Aberration corrections for free-space optical
communications in atmosphere turbulence using orbital angular momentum
states,\textquotedblright\ \textit{Opt. Express} \textbf{20}, 452--461 (2012).

\bibitem{Grilli01} S. Grilli, P. Ferraro, S. D. Nicola, A. Finizio, G.
Pierattini, and R. Meucci, \textquotedblleft Whole optical wavefields
reconstruction by digital holography, \textquotedblright\ \textit{Opt. Express} \textbf{9}, 294--302 (2001).

\bibitem{MigukinArXiv} A. Migukin, V. Katkovnik, and J. Astola,
\textquotedblleft Advanced phase retrieval: maximum likelihood technique
with sparse regularization of phase and amplitudear,\textquotedblright\
arXiv:1108.3251v1.

\bibitem{VK12-1} V. Katkovnik and J. Astola, \textquotedblleft High-accuracy
wave field reconstruction: decoupled inverse imaging with sparse modeling of
phase and amplitude,\textquotedblright\ \textit{J. Opt. Soc. Am. A} \textbf{29},
44--54 (2012).

\bibitem{VK12-2} V. Katkovnik and J. Astola, \textquotedblleft Phase
retrieval via spatial light modulator phase modulation in 4f optical setup:
numerical inverse imaging with sparse regularization for phase and
amplitude,\textquotedblright\ \textit{J. Opt. Soc. Am. A} \textbf{29}, 105--116
(2012).

\bibitem{MigukinSPIE12} A. Migukin, V. Katkovnik, and J. Astola,
\textquotedblleft Advanced multi-plane phase retrieval using Graphic
Processing Unit: augmented Lagrangian technique with sparse
regularization,\textquotedblright\ \textit{Proc. SPIE} \textbf{8429}, (2012).       

\bibitem{Jan} J. R. Magnus and H. Neudecker, \textit{Matrix Differential Calculus with Applications in Statistics and Econometrics}, 2nd ed., (Wiley, 1999).

\bibitem{Kreis} Th. Kreis, \textit{Handbook of Holographic Interferometry:
Optical and Digital Methods}, (Wiley-VCH, 2005).

\bibitem{Elad} M. Elad, \textit{Sparse and Redundant Representations: From
Theory to Applications in Signal and Image Processing} (Springer, 2010).

\bibitem{Han} D. Han, K. Kornelson, D. Larson, and E. Weber, \textit{Frames
for Undergraduates} (Student Mathematical Library, AMS, 2007).

\bibitem{Donoho} D. L. Donoho, \textquotedblleft Compressed
sensing,\textquotedblright\ \textit{IEEE Trans. Inf. Theory} \textbf{52},
1289--1306 (2006).

\bibitem{VK2011} V. Katkovnik, A. Danielyan, and K. Egiazarian,
\textquotedblleft Decoupled inverse and denoising for image deblurring:
variational BM3D-frame technique,\textquotedblright\ in \textit{Proceedings
of the International Conference on Image Processing} (ICIP), 3514 -- 3517
(2011).

\bibitem{Aram} A. Danielyan, V. Katkovnik, and K. Egiazarian,
\textquotedblleft Image deblurring by augmented Lagrangian with BM3D frame
prior,\textquotedblright\ in \textit{Workshop on Information Theoretic
Methods in Science and Engineering} (WITMSE), Tampere, Finland, (2010).

\bibitem{Aram12} A. Danielyan, V. Katkovnik, and K. Egiazarian,
\textquotedblleft BM3D frames and variational image
deblurring,\textquotedblright\ \textit{IEEE Trans. on Image Proc.} \textbf{21%
}, 1715--1728 (2012).

\bibitem{MigukinAL} A. Migukin, V. Katkovnik, and J. Astola, "Wave field
reconstruction from multiple plane intensity-only data: Augmented Lagrangian
algorithm", \textit{J. Opt. Soc. Am. A} \textbf{28}, 993--1002 (2011).

\bibitem{Bertsekas} D. P. Bertsekas, \textit{Nonlinear Programming}, 2nd ed.
(Athena Scientific, 1999).

\bibitem{Arizon} V. Arrizon, E. Carreon, and M. Testorf, \textquotedblleft
Implementation of Fourier array illuminators using pixelated SLM: efficiency
limitations,\textquotedblright\ \textit{Optics comm.} \textbf{16}, 207--213 (1999).

\bibitem{agour10} M. Agour, C. Falldorf, and C. von Kopylow,
\textquotedblleft Digital pre-filtering approach to improve optically
reconstructed wavefields in opto-electronic holography,\textquotedblright\ \textit{J. Opt.} \textbf{12}, 055401 (2010).
\newpage
\bibitem{Otsu} N. Otsu. \textquotedblleft A threshold selection method from
gray-level histograms,\textquotedblright\ \textit{IEEE Transactions of Systems, Man and Cybernetics} \textbf{9}, 62--66 (1979).

\bibitem{Agour} M. Agour, C. Falldorf, C. v. Kopylow, R. B. Bergmann,
\textquotedblleft Automated compensation of misalignment in phase retrieval
based on a spatial light modulator,\textquotedblright\ \textit{Appl. Opt.} \textbf{50}%
, 4779-4787 (2011).

\bibitem{AgourSPIE} M. Agour, C. Falldorf, C. von Kopylow and R. B.
Bregmenn, \textquotedblleft The effect of misalignment in phase retrieval
based on a spatial light modulator,\textquotedblright\ \textit{Proc. SPIE} \textbf{%
8082}, (2011).

\end{thebibliography}
\end{document}